\documentclass[twocolumn]{aastex63}

\usepackage{graphicx}
\usepackage{epstopdf}
\usepackage[intlimits,sumlimits]{amsmath}
\usepackage{epsfig}
\usepackage{amssymb}
\usepackage{mathrsfs}  
\usepackage{gensymb}

\usepackage{natbib}

\usepackage{amsmath}

\definecolor{dark-red}{rgb}{0.4,0.15,0.15}
\definecolor{dark-blue}{rgb}{0.15,0.15,0.4}
\definecolor{medium-blue}{rgb}{0,0,0.5}
\hypersetup{
    colorlinks, linkcolor={dark-red},
    citecolor={dark-blue}, urlcolor={medium-blue}
}

\newcommand{\beqa}{\begin{eqnarray}} 
\newcommand{\eeqa}{\end{eqnarray}}

\newcommand{\bsub}{\begin{subequations}}
\newcommand{\esub}{\end{subequations}}
\newcommand{\beal}{\begin{align}}
\newcommand{\ealn}{\end{align}}

\shorttitle{ZTF Bright Transient Survey Statistical Sample}

\begin{document}

\title{The Zwicky Transient Facility Bright Transient Survey.  II. \\ A Public Statistical Sample for Exploring Supernova Demographics\footnote{An interactive catalog with all data used in this paper is available at \url{https://sites.astro.caltech.edu/ztf/bts} and is updated in real time.}}

\correspondingauthor{Daniel~A.~Perley}
\email{d.a.perley@ljmu.ac.uk}

\author[0000-0001-8472-1996]{Daniel~A.~Perley}
\affiliation{Astrophysics Research Institute, Liverpool John Moores University, Liverpool Science Park, 146 Brownlow Hill, Liverpool L35RF, UK}

\author[0000-0002-4223-103X]{Christoffer~Fremling}
\affiliation{Division of Physics, Mathematics, and Astronomy, California Institute of Technology, Pasadena, CA 91125, USA}

\author[0000-0003-1546-6615]{Jesper~Sollerman}
\affiliation{Department of Astronomy, The Oskar Klein Center, Stockholm University, AlbaNova, SE-10691 Stockholm, Sweden}

\author[0000-0001-9515-478X]{Adam~A.~Miller}
\affiliation{Center for Interdisciplinary Exploration and Research in Astrophysics and Department of Physics and Astronomy, Northwestern University, 1800 Sherman Ave, Evanston, IL 60201, USA}
\affiliation{The Adler Planetarium, Chicago, IL 60605, USA}

\author{Aishwarya S. Dahiwale}
\affiliation{Division of Physics, Mathematics, and Astronomy, California Institute of Technology, Pasadena, CA 91125, USA}

\author[0000-0003-4531-1745]{Yashvi~Sharma}
\affiliation{Division of Physics, Mathematics, and Astronomy, California Institute of Technology, Pasadena, CA 91125, USA}

\author[0000-0001-8018-5348]{Eric C. Bellm}
\affiliation{DIRAC Institute, Department of Astronomy, University of Washington, 3910 15th Avenue NE, Seattle, WA 98195, USA}

\author[0000-0002-5741-7195]{Rahul~Biswas}
\affiliation{Department of Physics, The Oskar Klein Center, Stockholm University, AlbaNova, SE-10691 Stockholm, Sweden}

\author[0000-0001-5955-2502]{Thomas G. Brink}
\affiliation{Department of Astronomy, University of California, Berkeley, CA 94720-3411, USA}

\author[0000-0002-0786-7307]{Rachel J. Bruch}
\affil{Department of Particle Physics and Astrophysics, Weizmann Institute of Science, 234 Herzl St., 76100 Rehovot, Israel}

\author[0000-0002-8989-0542]{Kishalay De}
\affil{Division of Physics, Mathematics, and Astronomy, California Institute of Technology, Pasadena, CA 91125, USA}

\author[0000-0002-5884-7867]{Richard Dekany}
\affil{Caltech Optical Observatories, California Institute of Technology, Pasadena, CA  91125, USA}

\author{Andrew J. Drake}
\affiliation{Division of Physics, Mathematics, and Astronomy, California Institute of Technology, Pasadena, CA 91125, USA} 

\author[0000-0001-5060-8733]{Dmitry A. Duev}
\affiliation{Division of Physics, Mathematics, and Astronomy, California Institute of Technology, Pasadena, CA 91125, USA}

\author[0000-0003-3460-0103]{Alexei V. Filippenko}
\affiliation{Department of Astronomy, University of California, Berkeley, CA 94720-3411, USA}
\affiliation{Miller Senior Fellow, Miller Institute for Basic Research in Science, University of California, Berkeley, CA 94720 USA}

\author[0000-0002-3653-5598]{Avishay Gal-Yam}
\affiliation{Benoziyo Center for Astrophysics, The Weizmann Institute of Science, Rehovot 76100, Israel}

\author[0000-0002-4163-4996]{Ariel~Goobar}
\affiliation{Department of Physics, The Oskar Klein Center, Stockholm University, AlbaNova, SE-10691 Stockholm, Sweden}

\author{Matthew J. Graham}
\affiliation{Division of Physics, Mathematics, and Astronomy, California Institute of Technology, Pasadena, CA 91125, USA} 

\author[0000-0002-9154-3136]{Melissa L. Graham}
\affiliation{DIRAC Institute, Department of Astronomy, University of Washington, 3910 15th Avenue NE, Seattle, WA 98195, USA}

\author[0000-0002-9017-3567]{Anna Y. Q. Ho}
\affiliation{Division of Physics, Mathematics, and Astronomy, California Institute of Technology, Pasadena, CA 91125, USA}
\affiliation{Department of Astronomy, University of California, Berkeley, CA 94720-3411, USA}
\affiliation{Miller Institute for Basic Research in Science, University of California, Berkeley, CA 94720, USA}

\author[0000-0002-7996-8780]{Ido Irani}
\affiliation{Benoziyo Center for Astrophysics, The Weizmann Institute of Science, Rehovot 76100, Israel}

\author[0000-0002-5619-4938]{Mansi M. Kasliwal}
\affiliation{Division of Physics, Mathematics, and Astronomy, California Institute of Technology, Pasadena, CA 91125, USA}

\author[0000-0002-1031-0796]{Young-Lo Kim}
\affiliation{Universit\'e de Lyon, Universit\'e Claude Bernard Lyon 1, CNRS/IN2P3, IP2I Lyon, F-69622, Villeurbanne, France}

\author[0000-0001-5390-8563]{S. R. Kulkarni}
\affiliation{Division of Physics, Mathematics, and Astronomy, California Institute of Technology, Pasadena, CA 91125, USA} 

\author{Ashish Mahabal}
\affiliation{Division of Physics, Mathematics, and Astronomy, California Institute of Technology, Pasadena, CA 91125, USA}
\affiliation{Center for Data Driven Discovery, California Institute of Technology, Pasadena, CA 91125, USA}

\author[0000-0002-8532-9395]{Frank J. Masci}
\affiliation{IPAC, California Institute of Technology, 1200 E. California Blvd, Pasadena, CA 91125, USA}

\author[0000-0002-8532-827X]{Shaunak Modak}
\affiliation{Department of Astronomy, University of California, Berkeley, CA 94720-3411, USA}

\author[0000-0002-0466-1119]{James D. Neill}
\affiliation{Division of Physics, Mathematics, and Astronomy, California Institute of Technology, Pasadena, CA 91125, USA}

\author[0000-0001-8342-6274]{Jakob Nordin}
\affiliation{Institut f{\"u}r Physik, Humboldt-Universit{\"a}t zu Berlin, Newtonstr. 15, 12489 Berlin, Germany}

\author[0000-0002-0387-370X]{Reed L. Riddle}
\affiliation{Caltech Optical Observatories, California Institute of Technology, Pasadena, CA 91125, USA}

\author[0000-0001-6753-1488]{Maayane T. Soumagnac}
\affiliation{Lawrence Berkeley National Laboratory, 1 Cyclotron Road, Berkeley, CA 94720, USA}
\affiliation{Department of Particle Physics and Astrophysics, Weizmann Institute of Science, Rehovot 76100, Israel}

\author[0000-0002-4667-6730]{Nora L. Strotjohann}
\affiliation{Benoziyo Center for Astrophysics, The Weizmann Institute of Science, Rehovot 76100, Israel}

\author[0000-0001-6797-1889]{Steve Schulze}
\affiliation{Benoziyo Center for Astrophysics, The Weizmann Institute of Science, Rehovot 76100, Israel}

\author[0000-0001-8472-1996]{Kirsty~Taggart}
\affiliation{Astrophysics Research Institute, Liverpool John Moores University, Liverpool Science Park, 146 Brownlow Hill, Liverpool L35RF, UK}

\author[0000-0003-0484-3331]{Anastasios Tzanidakis}
\affiliation{Division of Physics, Mathematics, and Astronomy, California Institute of Technology, Pasadena, CA 91125, USA} 

\author[0000-0001-8018-5348]{Richard S. Walters}
\affiliation{Division of Physics, Mathematics, and Astronomy, California Institute of Technology, Pasadena, CA 91125, USA} 

\author[0000-0003-1710-9339]{Lin Yan}
\affiliation{Caltech Optical Observatories, California Institute of Technology, Pasadena, CA 91125, USA}

\begin{abstract}
We present a public catalog of transients from the Zwicky Transient Facility (ZTF) Bright Transient Survey (BTS), a magnitude-limited ($m<19$ mag in either the $g$ or $r$ filter) survey for extragalactic transients in the ZTF public stream.  We introduce cuts on survey coverage, sky visibility around peak light, and other properties unconnected to the nature of the transient, and show that the resulting statistical sample is spectroscopically 97\% complete at $<$18 mag, 93\% complete at $<$18.5 mag, and 75\% complete at $<$19 mag.   We summarize the fundamental properties of this population, identifying distinct duration-luminosity correlations in a variety of supernova (SN) classes and associating the majority of fast optical transients with well-established spectroscopic SN types (primarily SN Ibn and II/IIb).  We measure the Type Ia SN and core-collapse (CC) SN rates and luminosity functions, which show good consistency with recent work.  About 7\% of CC~SNe explode in very low-luminosity galaxies ($M_i>-16$ mag), 10\% in red-sequence galaxies, and 1\% in massive ellipticals.  We find no significant difference in the luminosity or color distributions between the host galaxies of Type II and Type Ib/c supernovae, suggesting that line-driven wind stripping does not play a major role in the loss of the hydrogen envelope from their progenitors.  Future large-scale classification efforts with ZTF and other wide-area surveys will provide high-quality measurements of the rates, properties, and environments of all known types of optical transients and limits on the existence of theoretically predicted but as of yet unobserved explosions.
\end{abstract}


\section{Introduction} \label{sec:intro}

Recent years have brought an unprecedented expansion in our ability to survey for transient astronomical phenomena. 
The optical sky is now being scanned on an almost nightly basis by several different telescope networks around the world, including the All-Sky Automated Survey for Supernovae \citep[ASAS-SN;][]{Shappee2014}, the Asteroid Terrestrial Last-Alert System \citep[ATLAS;][]{Tonry2018}, and the Zwicky Transient Facility \citep[ZTF;][]{Bellm2019,Bellm2019b,Graham2019,Dekany20}. New projects, such as BlackGEM \citep{Groot2019} and the Gravitational-wave Optical Transient Observer \citep[GOTO;][]{Dyer2018} are  
beginning operations, and earlier surveys such as the Panoramic Survey Telescope and Rapid Response System \citep[Pan\nobreakdash-STARRS;][]{Kaiser2002} continue.  In 2019 almost twenty thousand new and unique optical transients were reported via official channels \citep{Kulkarni2020}, an increase of two orders of magnitude from a decade prior \citep{GalYam2013}.  The Legacy Survey of Space and Time (LSST; \citealp{Ivezic2019}) at the Vera C. Rubin Observatory is expected to increase these numbers by another order of magnitude within a few years.

Large numbers of transients are of limited scientific value without secure classifications and redshifts \citep{Kulkarni2020}.  Despite recent advances in photometric classification (e.g., \citealp{Muthukrishna+2019,Villar+2019,Villar+2020,Dauphin2020,Hosseinzadeh+2020}), the only ground truth for this remains spectroscopy, an observationally expensive endeavor.  Deciding which transients to spectroscopically classify and which to ignore typically involves extensive human decision-making, potentially introducing complex biases and diminishing the value of large statistical samples for studies of (for example) volumetric rates, luminosity functions, or ensemble host-galaxy properties.

In \cite{Fremling2020} we introduced the ZTF Bright Transient Survey (BTS), which aims to provide a large and purely magnitude-limited ($m<19$ mag for discovery and $m<18.5$ mag for classification\footnote{A parallel volume-limited survey, the ZTF Census of the Local Universe (CLU) experiment, extends the classification threshold to $m<20$ mag for transients occurring in known galaxies within $D<200$ Mpc \citep{De+2019,De2020CLU}.}) sample of extragalactic transients in the northern sky, suitable for detailed statistical and demographic analysis.  In that work we described some of the aims of the project and presented early results on the fraction of supernovae (SNe) at this magnitude level hosted by galaxies with known, cataloged redshift (44\%), along with a catalog of the first 761 SNe found by the project.

The BTS is an ongoing effort that will continue in its current form through the end of the public ZTF Northern Sky Survey \citep{Bellm2019b} in October 2020.  Work is ongoing to provide final photometric and spectroscopic data releases (and associated scientific papers) spanning this entire period.  Preliminary photometric and spectroscopic data are also released in real time on a nightly basis via the ZTF brokers\footnote{Currently operating, fully-featured public brokers include ANTARES (\url{https://antares.noao.edu/}), LASAIR (\url{https://lasair.roe.ac.uk/}), ALERCE (\url{http://alerce.science/}), and MARS (\url{https://mars.lco.global/})} and the Transient Name Server (TNS\footnote{\url{https://wis-tns.weizmann.ac.il}}).  In this paper, which is accompanied by an online web portal, we supplement these basic, continuous data releases with a live catalog of higher-level properties of our sample measured from the real-time public data --- in particular, peak luminosities, rise and decay times, and host-galaxy associations --- and demonstrate the use of the sample for a variety of scientific aims.

The paper is structured as follows.  In \S\,\ref{sec:selection} we describe additional improvements to the filtering and screening process implemented since 2018, and detail a series of post-facto selection cuts that we employ to remove poorly-observed transients and variables without imposing selection biases on the remaining sample. \S\,\ref{sec:completeness} provides spectroscopic completeness statistics for the resulting subset, demonstrating that it is $\sim93$\% complete for transients peaking above $m<18.5$ mag.  In \S\,\ref{sec:results} we highlight how in only two years ZTF has mapped out a vast swathe of the observational transient parameter space, providing the largest and most reliable look at the diversity of luminous transient phenomena in the Universe.   We also provide preliminary characterizations of the SN luminosity function, the core-collapse SN rate, and a color-magnitude analysis of host-galaxy properties of the major SN classes.  We summarize our work in \S\,\ref{sec:conclusions} and provide additional documentation of the BTS Sample Explorer, a public webpage which serves our real-time transient catalog.

\section{BTS Sample Selection and Characterization}
\label{sec:selection}

\begin{figure*}
\centering
\includegraphics[width=17cm]{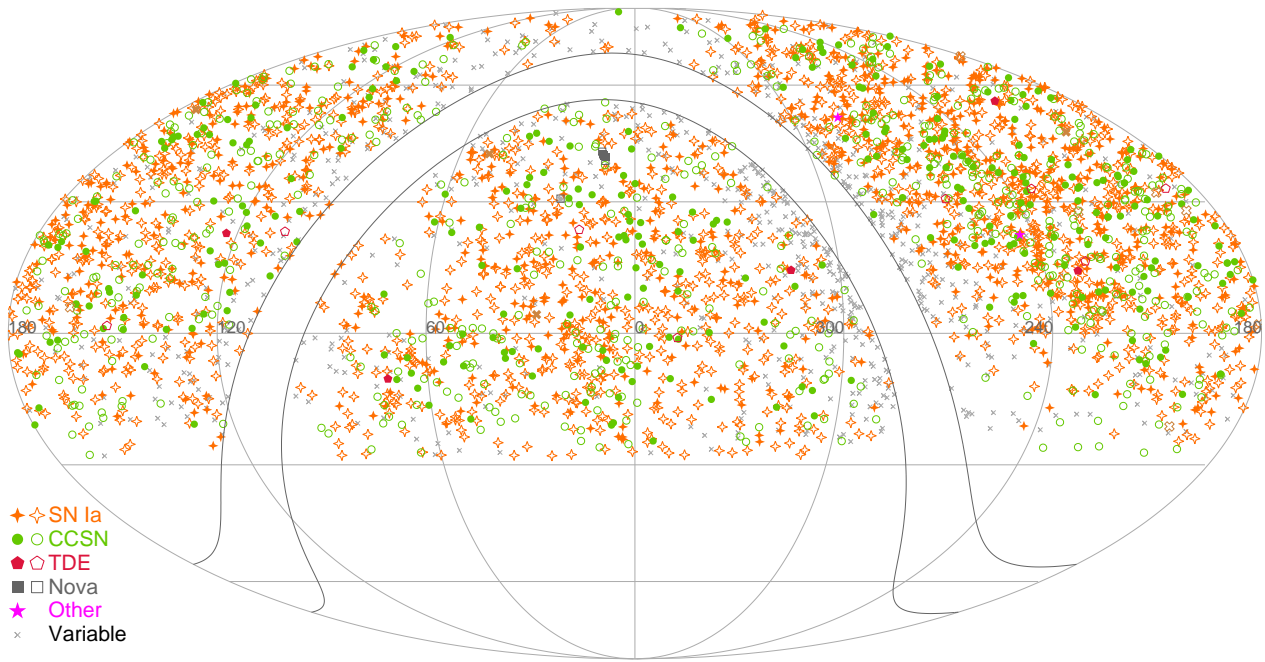}
\caption{Skymap of classified transients and variables within BTS.  Transients at $m_{\rm peak}<18.5$ mag satisfying our sample cuts (described in \S\,\ref{sec:quality} and \S\,\ref{sec:purity}) are shown as solid symbols; other transients are shown as open symbols.  BTS covers the entire northern sky outside the Galactic plane, although there is some variation in transient sky-areal density as a result of seasonal variations in survey coverage.}
\label{fig:skymap}
\end{figure*}

\subsection{Alert Filter}
\label{sec:filter}

The fundamental criteria for inclusion of an event in the BTS are that it is a genuine transient (a well-defined \emph{event} with a beginning and end, as distinct from a variable star or active galactic nucleus [AGN] for which changes in flux are always occurring), that it is extragalactic, and that it is brighter than $m<19$ mag in either the $g$ or $r$ filters at some point in its observed ZTF light curve.  
On any given night the number of genuine transients at this magnitude level within the ZTF difference-image alert stream \citep{Masci2019,Patterson2019} is vastly outnumbered by variables (stars and AGNs), artifacts, and moving objects.  These must be efficiently filtered out without losing any of the genuine transients the survey seeks to catalog.

All ZTF Avro\footnote{\url{https://avro.apache.org}} packets contain two machine-learning scores to aid this process: a real-bogus score (\texttt{rbscore}; \citealp{Mahabal2019}) to separate PSF-like sources from artifacts in ZTF subtractions, and a star-galaxy separation score (\texttt{sgscore}; \citealt{Tachibana2018}), based on a cross-match with the Pan-STARRS \citep{Chambers2016} catalog, to aid in the rejection of stars.  Neither metric is perfect, and producing a transient catalog free of variables and artifacts requires substantial additional effort using both software filters and human attention.

Our first-year in-stream software filter, which reduced the $\sim10^6$ Avro alert packets produced each night to $\sim500$ viable transient candidates, was first described by \cite{Fremling2020}.  The cuts employed by this filter were relatively basic: two detections at least 0.02 days ($\sim30$ min) apart, a high \texttt{rbscore}, no underlying counterpart (${<}\,2\arcsec$) with high \texttt{sgscore}, no bright star in the vicinity (${<}\,20\arcsec$), and a difference magnitude brighter than $m<19$.  While this filter had the benefit of being straightforward to implement and understand, the false-positive rate was significant (several hundred stars/AGNs would pass the filter nightly) and on rare occasions it would miss SNe near bright galaxy nuclei mistakenly flagged as stars by \texttt{sgscore}.  Thus, beginning in June 2019 we have made several additional adjustments.

Events with a long history of previous detections coincident with a bright PS1 or Gaia source are now rejected, since these tend to be AGNs flagged as galaxies.  The exclusion radius around bright stars has been reduced (to an extent depending on the star's brightness and color) to reduce the risk of rejecting objects around galaxies mistakenly flagged as stars.  We also remove slow-moving asteroids using a catalog cross-match and employ the new deep-learning real-bogus \texttt{drb} score \citep{Duev2019} to better remove artifacts.  A full list of changes is given in Appendix~\ref{sec:filterdetail}. 
These adjustments reduced the typical number of false positives from several hundred per night to $\lesssim50$, reducing (but not entirely eliminating) the need for human vetting.

The BTS filter runs in parallel on the GROWTH Marshal \citep{Kasliwal2019} and on \texttt{AMPEL} \citep{Nordin2019}.  Candidates passing the filter {(and not already saved to the BTS program)} are reviewed by human scanners nightly {using the GROWTH Marshal scanning tool, as described by \cite{Fremling2020}}.  Candidates assessed to definitely be AGNs, variable stars, artifacts, or other false positives are ignored; the remaining candidates are registered to the BTS program within the GROWTH Marshal database (``saved'').  {Typically 5--10 candidates are saved on an average clear night.}  A skymap of saved events with classifications is shown in Figure~\ref{fig:skymap}.

\subsection{Candidate Characterization}
\label{sec:charac}

Once saved, candidates are subjected to additional scrutiny: visual inspection of the full alert-based light curve and cross-matches to various catalogs and imaging data.  This is sometimes enough to classify false positives (for example, if the transient is coincident with a known AGN or a WISE source with AGN-like colors and the light curve shows normal AGN variability, or if a faint star-like counterpart is visible in the image and previous flares from this location are seen in the ZTF light curve or reported on TNS).  Otherwise, the candidate is reported to TNS and spectroscopic follow-up observations with the Spectral Energy Distribution Machine (SEDM; \citealt{Blagorodnova2018,Rigault2019}) on the Palomar 60-inch telescope or the Spectrograph for the Rapid Acquisition of Transients (SPRAT; \citealt{SPRAT}) on the Liverpool Telescope \citep{Steele+2004} is requested according to our priority-ranking system \citep{Fremling2020} and reviewed on an approximately weekly basis.  Targets that cannot be classified with these facilities are scheduled for observations during scheduled classical observing runs at larger telescopes --- in particular with the Double-Beam Spectrograph (DBSP; \citealt{Oke1982}) at the Palomar 5~m Hale telescope, with the Dual Imaging Spectrograph (DIS) at Apache Point Observatory, the Kast spectrograph on the Shane 3~m telescope at Lick Observatory \citep{Miller-Stone-1993}, and occasionally with the Low-Resolution Imaging Spectrometer (LRIS; \citealt{Oke1995}) at the 10~m Keck-I telescope on Maunakea.  Spectra are reduced and resulting classifications are registered with TNS, generally within 24 hr of observations.  If a classification is first reported to TNS by another group or acquired by another ZTF program unrelated to our project, we avoid duplicating effort unless there is an indication that the external classification is unreliable.  (See \citealt{Fremling2020} for additional details and statistics on classifications from the first year of the project.)

\begin{figure*}
\centering
\includegraphics[width=18cm]{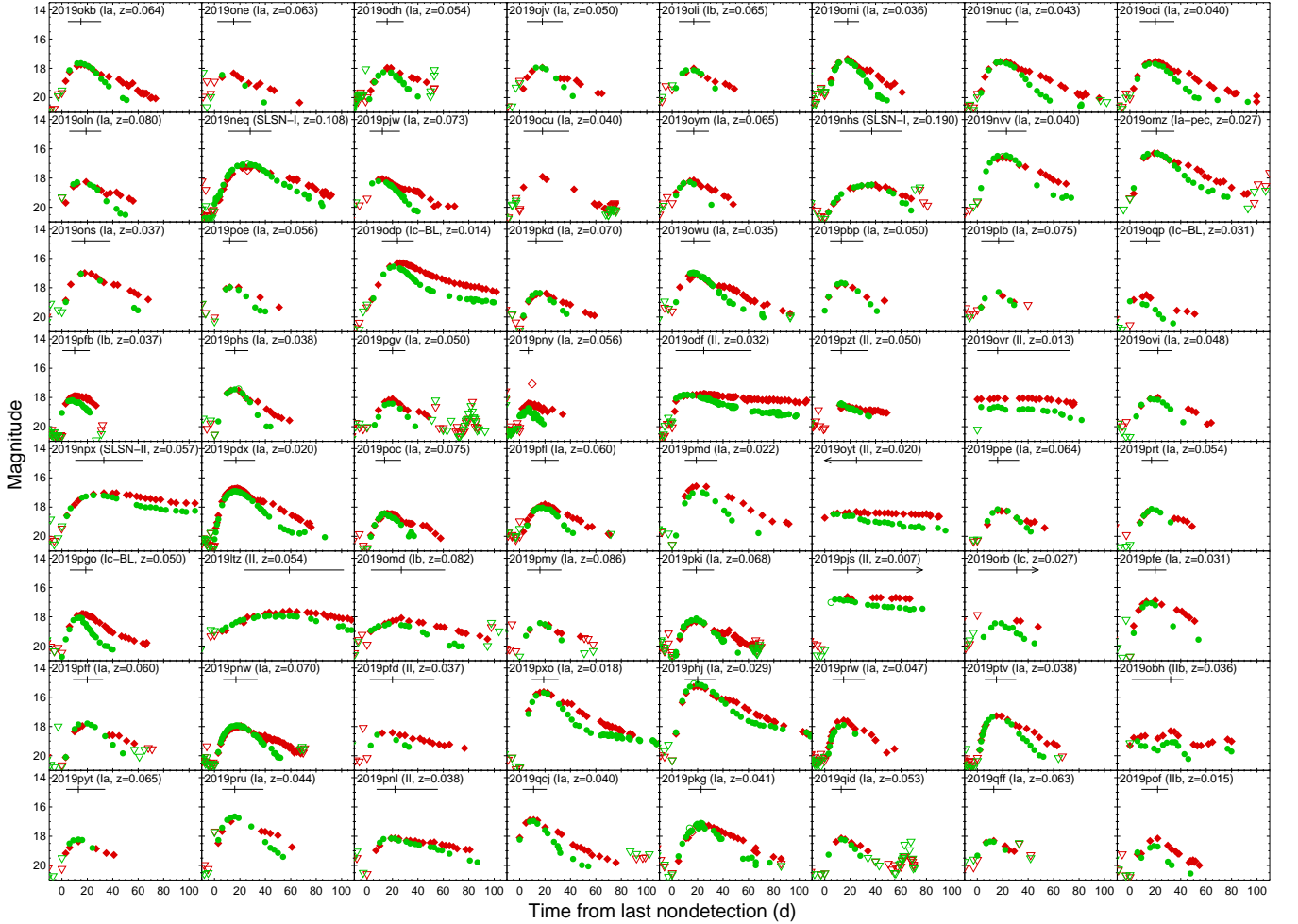}
\caption{Example two-filter ZTF light curves for all events that passed our quality cuts and which peaked at $m<18.5$ mag during a 22-day period in summer 2019 (JD 2458727.8--2458749.8).  Green circles are $g$ band and red diamonds are $r$ band.  Open triangles signify upper limits and unfilled circles/diamonds are low-quality measurements not used in the light curve measurements.  
The cross symbol at upper left of each panel shows the rise, peak, and fade times (the left end, crossbar, and right end of the cross, respectively) measured using the technique as defined in \S\,\ref{sec:charac}.}
\label{fig:lightcurves}
\end{figure*}

All saved candidates are also automatically analyzed by an independent script that continuously downloads the public Avro packet data for every transient in our program.  Light curves are built from the packet data using the \texttt{jd} {(time)}, \texttt{fid} {(filter)}, \texttt{magpsf} {(magnitude)}, \texttt{sigmapsf} {(uncertainty)}, and \texttt{diffmaglim} {(field limiting magnitude)} values; examples are shown in Figure~\ref{fig:lightcurves}.  Measurements in poor observing conditions (indicated by limiting magnitudes shallower than 19) are ignored.   For both the $g$-band and $r$-band light curves, we measure the time (JD) and magnitude of observed maximum light as the brightest measurement in the relevant light curve.  The ``rise time'' and ``fade time'' are calculated as the time elapsed in days between this peak and the point where the light curve drops to 0.75 mag below peak (equivalent to half the peak flux); this is calculated using simple linear interpolation between data points.  Upper limits from the alert-packet history are used only if they occur before the first detection, in which case the interpolation is performed between the limit and the first detection in the light curve.\footnote{Limits in packet data refer to the limiting magnitude for empty regions of the whole ZTF field, and only designate that an alert was not generated.  This usually means a true nondetection, but can also indicate a data-quality issue, leading to apparent nondetections occasionally being interspersed throughout a light curve.}
This calculation is run separately in each filter, but because filter coverage is often irregular (e.g., the rise may be sampled in the $g$ band but the decline only in the $r$ band) we also calculate a hybrid timescale measurement by shifting the ``fainter'' band (as defined at peak) to match the ``brighter'' one at its peak, and using whichever band has the fastest evolution in each direction.  (This hybrid measurement is the value used in the cuts and subsequent analysis discussed later.)

We also obtain the Galactic extinction along the line of sight using the NASA Extragalactic Database (NED) extinction tool\footnote{\url{https://ned.ipac.caltech.edu/extinction_calculator}} (based on the dust map of \citealp{Schlafly11}).  If the redshift of the event has been measured, we calculate the peak (observed) absolute magnitude by applying the extinction correction, the distance modulus calculated from the redshift assuming a cosmological model with $\Omega_M=0.3$, $\Omega_\Lambda=0.7$, and $h=0.7$ (peculiar-motion corrections are ignored, although for transients in M31 and M33 we use standard distance moduli of 24.4 and 24.55 mag, respectively), and apply a uniform $K$-correction of $2.5 \, {\rm log}_{10}(1+z)$.  
This estimate will be close to the real absolute magnitude for transients passing our quality cuts and with characteristic timescales longer than several days, but is a lower limit for transients for which the peak was not sampled or very fast events.   Note also that redshifts measured from SN features have significant uncertainty ($\Delta z \approx 0.005$; \citealt{Fremling2020}) and the impact of peculiar motions (up to $\Delta v \approx 600$ km s$^{-1}$, or $\Delta z \approx 0.002$) can also be significant at low redshift, so there is additional scatter in these measurements for nearby transients (up to 0.5 mag at $z \approx 0.02$ if the redshift is not precisely known, or 0.2 mag if a host-galaxy redshift is available).  About 6\% of our classified sample is at $z<0.02$; while in the majority of these cases ($\sim75$\%; see also \citealt{Fremling2020}) precise host-galaxy redshifts are available, this uncertainty should be kept in mind when dealing with low-luminosity populations or in interpreting luminosity outliers.

We additionally cross-match the location of the transient to a variety of catalogs.  We use the nearest three PS1 cross-matches from the Avro packets (which contain \texttt{sgscore} values).  We also download the list of potential cross-matches registered on the event summary page on Lasair \citep{Smith+2019}.  Lasair also provides star/galaxy classifications,  although we employ this information only for objects brighter than $g<21$ or $r<21$ mag.  If a cataloged star exists within 1$\arcsec$ of the candidate we associate it with the (probable) star.  Otherwise, we calculate the most likely galaxy counterpart using a chance-probability calculation: we exclude stars and choose the cross-match for which the probability of finding a galaxy as bright (or brighter) as close (or closer) to the transient position by random chance is lowest (Appendix~\ref{sec:xmatchprob}).  This is done independently for both cross-match lists and the results are used for our purity cut (\S\,\ref{sec:purity}).

To provide additional counterpart photometry for host-galaxy analysis purposes, we also download the complete Sloan Digital Sky Survey (SDSS) and Pan-STARRS (PS1) photometric catalogs for sources near the transient location and calculate the most likely cross-match in this region.  Stars and galaxies are distinguished using the SDSS star-galaxy code, or the PS1 PSF-Kron magnitude method\footnote{\url{https://outerspace.stsci.edu/display/PANSTARRS/How+to+separate+stars+and+galaxies}}.  The method for identifying the probable counterpart association is similar to the Avro/Lasair cross-match method above, although we also use the transient classification to restrict the search to exclude matches of inappropriate type (e.g., potential stellar associations are excluded for classified SNe).  {The cross-match is based only on the SDSS/PS1 photometry; spectroscopic information or other cross-matched catalogs are not yet considered.}  The search is currently limited to within $<90\arcsec$ and $<30$ kpc (projected) of the transient location to minimize false positives, although these restrictions will be loosened in the future as more catalogs are added.  If there is no source meeting these criteria the event is designated as hostless and an upper limit is calculated using the relevant survey limiting magnitudes.

\subsection{Quality Filter: Removal of Poorly-Observed Candidates}
\label{sec:quality}

Our general aim is to acquire spectra of every event which \emph{could} be a SN and which exceeds $m \leq 18.5$ mag at any point in its evolution when observed by the survey, regardless of any other properties of the object.  However, owing to poor weather, seasonal gaps, or a suboptimal sky location at the time of discovery, some events may be especially difficult to characterize even if they are bright.

These limitations are unavoidable for an all-sky, ground-based survey, but their impact on the completeness and quality of our survey can be minimized by introducing additional, post-facto cuts to remove events discovered at times and sky locations that were heavily affected by coverage gaps or poor observability.  Care must be taken that these cuts are unbiased with respect to the properties of the transient to the maximum extent possible.  In particular, they should be minimally dependent on duration: for example, requiring a large number of detections would introduce a strong bias against short-timescale transients.  

Our current set of quality cuts is as follows.  

\begin{enumerate}
    \item The transient must have an observation significantly prior to peak light.  Specifically, P48 must have observed the field at least once between 7.5 and 16.5 days prior to the time when the brightest detection in the light curve was recorded.  The transient need not be detected in this observation, but the observation must be deep enough to be constraining ($m_{\rm lim} > 19$ mag).
    \item The transient must have two observations around the time of peak light.  Specifically, in addition to the observation at (apparent) maximum, the transient must have a second observation either 2.5 to 7.5 days before \emph{or} 2.5 to 7.5 days after this measurement.  
    These measurements must also have $m_{\rm lim} > 19$ mag.
    \item The transient must have an observation after peak light. Specifically, it must have an observation between 7.5 and 16.5 days after the observed time of maximum, or alternatively an observation 2.5 to 7.5 days after maximum \emph{and} an observation between 16.5 and 28.5 days after maximum.
    These measurements must also have $m_{\rm lim} > 19$ mag.
    \item The location of the transient in the sky must be conducive to follow-up spectroscopy. Specifically, it must remain above 30 degrees elevation for at least 2 hr at $>12^\circ$ twilight during the night occurring 30 days after the observed time of peak light.
    \item The transient must not be present in its reference image.  Specifically, there must be at least a 30-day span between the last exposure in its reference image and the first registered detection in its light curve.
    \item Galactic extinction toward the transient should be low ($A_V < 1.0$ mag).
    \item At least one packet in the alert history must pass the most recent version of the alert-stream filter (\ref{sec:filter}), even if the candidate was saved under an earlier version of the filter.
\end{enumerate}

{Cuts 1--3 limit the sample to events for which the peak time and luminosity are well-constrained, and approximately weekly cadence is maintained between two weeks prior to peak and two weeks after peak.  Cuts 4--6 remove events that present other types of difficulties (e.g., hard to obtain spectra, photometry is contaminated, large/uncertain extinction correction).  The final cut is not strictly a quality cut but is applied for consistency.}  For the purposes of this paper, we additionally restrict the sample to transients with a time of peak between 2018-06-01 (the public start of the BTS survey) and 2020-07-15, inclusive.

We emphasize that these criteria make no reference to the timescale or behavior of the transient itself.  The only assumption is that a well-defined single peak does exist and can be recognized based on three or more observations spanning two weeks around this peak.  
Transients with durations shorter than the permissible coverage gaps (about 10 days) will be mildly\footnote{We use the term ``mildly'' selected against to refer to types of events that may be undercounted by a factor of up to two, versus ``heavily'' selected against events which may be undercounted by a factor of more than two.  Exact selection losses will depend sensitively on light-curve shape and peak magnitude as well as on duration.   A detailed quantification of selection losses in these regimes is beyond the scope of this paper but will be addressed in future work using simulations.} selected against and transients with durations shorter than the survey cadence itself (3 days) will be heavily selected against.  Very slow transients with durations comparable to the survey (2 yr to date) will also be preferentially missed.   However, events with durations between $\sim 10$ and 100 days should be selected with equal efficiency as long as they have only a single peak.

Multi-peaked events or events with an extended, flat plateau are somewhat \emph{more} likely to be selected than single-peaked events, since they have multiple opportunities for the peak to fall in a window that passes our criteria.  These events are relatively rare, so this effect is not large, although a detailed measurement of duration-dependent rates would require additional corrections.

Together these cuts remove about half (48\%) of the sample.  About 25\% of the down-selection can be attributed to weather and instrument-downtime gaps, 12\% to other coverage gaps (lunar, seasonal, or scheduling), 5\% to visibility, 5\% to reference imaging, and 1\% to Galactic extinction.  Significant losses of this type are unavoidable, given our goal to establish a universal set of criteria for demographic studies of transients spanning a wide variety of potential behaviors.  Studies focused on a narrow range of events would be able to achieve a higher yield, at the expense of this generality, by employing different criteria tuned to the anticipated properties of the sample of interest.

\subsection{Purity Filter: Removal of False Positives}
\label{sec:purity}

It is inevitable that many candidates saved by the human scanners eventually turn out to be variables, rather than genuine transients: typically, AGNs or cataclysmic variable stars (CVs).  While these are not part of our project, they do pass the filter routinely and cannot always be distinguished from a transient given the information available at the time of scanning.  
Some are observed and classified spectroscopically by our observing programs, but whenever possible photometric and contextual information is used to preserve scarce spectroscopic resources for genuine extragalactic transients.

AGNs can often be recognized immediately after being saved via cross-matches to pre-existing spectroscopic surveys or to multiwavelength (radio, X-ray, mid-IR) catalogs.  Alternatively, it is possible to eliminate them photometrically after further monitoring: continuous slow variability (both upward and downward) lasting more than a year---with no evidence of a return to the pre-detection flux seen in the reference image---is also considered sufficient for a photometric-only classification.

CV eruptions (dwarf novae) usually exhibit a distinctive fast rise to peak followed by a somewhat slower decay during which a constant, blue $g-r$ color is maintained, often followed by a sudden drop; no known class of SN shows this behavior.  Dwarf novae are also expected to have a star-like counterpart, or no detectable counterpart, in PS1 reference imaging.  Candidates matching all of these criteria are considered to be securely classified as dwarf novae and are removed from the sample, even in the absence of spectroscopy.  Repeated flares (including flares in other surveys) separated by long stretches of inactivity are an even more definitive indicator of a CV origin.  

However, it is not always possible to obtain a definitive assessment of a false positive from the light curve alone: the short durations of dwarf novae in particular make it easy to miss one or more of the key phases above during brief gaps in coverage, introducing a potential to be confused on an individual basis with fast-evolving SNe.  Given our goal to produce a purely magnitude-limited, unbiased, and highly complete transient sample, it is still desirable to separate these from genuine transients in a systematic way.

\begin{figure*}
\centering
\includegraphics[width=18cm]{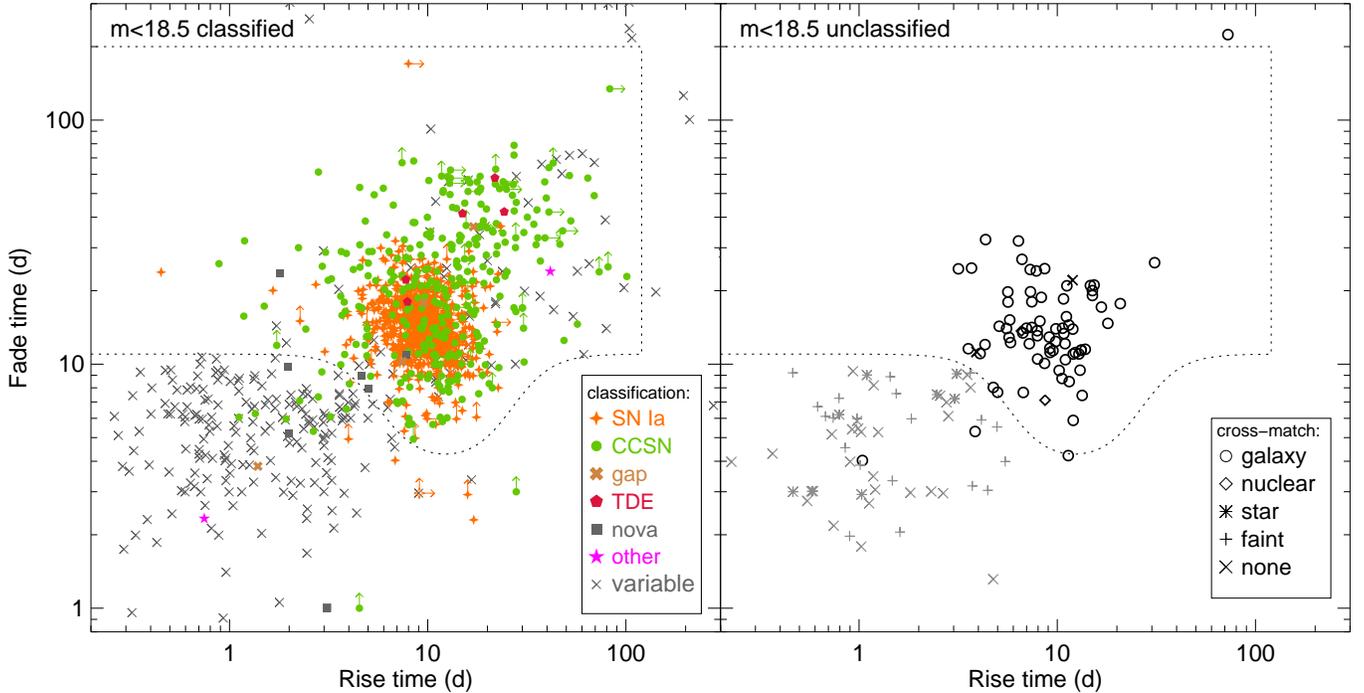}
\caption{Rise time (half-peak to peak) versus fade time (peak to half-peak) for $m_{\rm peak,obs} < 18.5$ mag events in BTS that pass the quality filter.  Events with public classifications (which are always spectroscopic for SNe/transients) are shown in the left panel.  Transients and variables cluster in different regions of the diagram, with some overlap.  The right panel shows events we were unsuccessful at classifying, which exhibit a similar bimodal distribution.  The symbol indicates the nature of the cross-matched catalog object.  Events that \emph{either} pass the light-curve cuts indicated by the dashed line \emph{or} have a credible cross-matched galaxy and are not coincident with its nucleus pass our purity cut.  At $m<18.5$ mag, 93\% of such events are classified and only 7\% are unclassified.  A small number of outlier SNe Ia with apparent very fast rise/fade times are present. This is usually due to erroneous upper limits associated with the failure of the transient to generate an alert, but can also be due to additional photometric scatter from subtraction residuals in bright, point-like galaxy nuclei.}
\label{fig:risefade}
\end{figure*}

To remove highly-probable false positives, we apply an additional criterion (which we will refer to as a purity cut): the potential transient must \emph{either} be coincident with a cross-matched galaxy (but not with its nucleus), \emph{or} its light curve must have a vaguely SN-like timescale.  The cross-match criterion is documented in Appendix~\ref{sec:xmatch}.  To satisfy the light-curve criterion, the rise time (see \S\,\ref{sec:charac}) must be $<120$ days and the fade time must be $<200$ days but more than $\sim 11$ days\footnote{The fade-time limit is slightly less stringent if the rise time is in the range of normal SNe. The exact equation for the fade-time lower limit is $t_{\rm fade} > 4 + 7/(1+e^{({\rm log}_{10}(5)+{\rm log}_{10}({t_{\rm rise}}))/0.1}) + 7/(1+e^{{\rm log}_{10}(30)-{\rm log}_{10}(t_{\rm rise}))/0.1})$}  (the region within the dashed line in Figure~\ref{fig:risefade}).  This selection in principle imposes a small bias against very fast transients in undetected galaxies or very compact galaxies (which mimic CVs), or against extremely slow transients in galaxy nuclei (which mimic AGNs). However, it is essentially unbiased to the parameter space occupied by the vast majority of real extragalactic transients.

The impact of our purity cut is visualized in Figure~\ref{fig:risefade}.  Nearly all spectroscopic SNe have SN-like light-curve properties by our definition (although there are a few exceptions).  Nearly all unclassified events with SN-like properties are associated with galaxies.  Since we only require that a candidate meets one of these criteria, this means that it will be very rare for a genuine transient to fail the purity cut, presuming its light curve is well sampled and free of bad measurements.

In practice, among events with classifications (and which pass the quality cuts), the purity cut removed only two classified transients ($<0.2$\% of the sample): one SN for which the timescale measurements were compromised by sparse data, and one (probable) nova in the far outskirts of M31 with an erratically flaring light curve.  Notably, none of the known fast transients failed the purity cut because all are associated with galaxies.  We also visually inspected all unclassified $m<18.5$ mag events that passed the quality cuts but failed the purity cut, and confirmed that none of them is likely to be a SN or some other extragalactic transient.

\section{Sample Completeness}
\label{sec:completeness}

The BTS project registered 3147 spectroscopically-classified transients during the 25.5-month period of this study, 1865 of which satisfy our quality and purity cuts (a partial breakdown of these by type is given in Table~\ref{tab:sncount}).
Given the stated goal of our project to provide a spectroscopically-complete magnitude-limited sample of extragalactic transients, we also need to know how many genuine transients met our selection criteria but could not be classified.

\begin{figure}
\centering
\includegraphics[width=8.6cm]{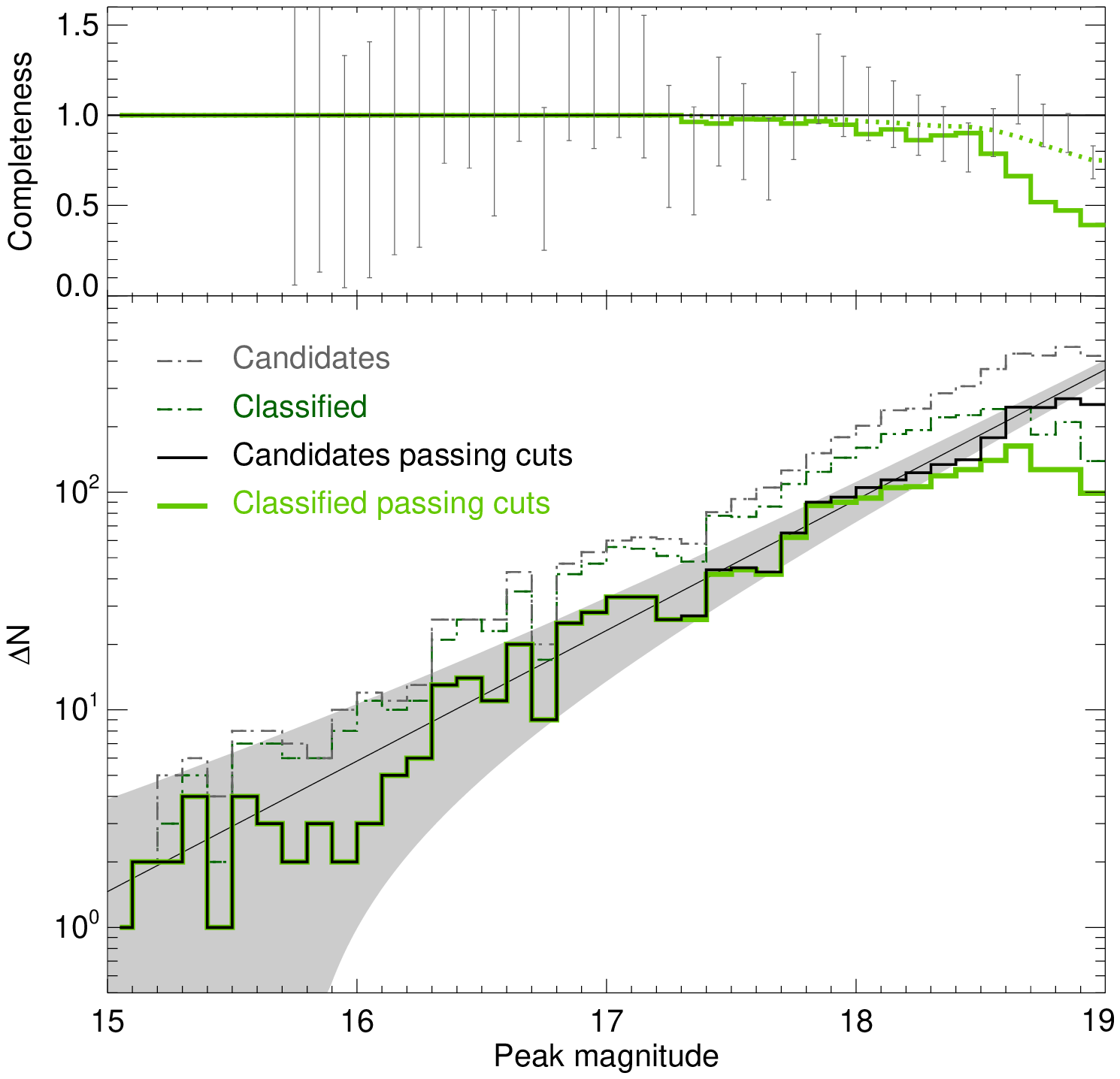}
\caption{Histograms of transient candidate and confirmed transient counts by magnitude.  The dashed histogram lines show all events saved to the survey not known to be variables.  The solid histogram lines show only events passing our sample cuts.   A fit assuming a simple $\frac{{\rm d}N}{{\rm d}\,{\rm log}f} \propto f^{3/2}$ power law, with 2$\sigma$ prediction intervals for each magnitude bin, is also illustrated.  Completeness fractions are shown in the upper panel.  Error bars indicate recovery completeness for saved sample transients with respect to the $f^{3/2}$ prediction (95\% Poisson confidence interval); the green line indicates spectroscopic completeness for saved transients passing sample cuts (the solid line shows completeness per bin, the dotted line shows cumulative completeness down to a particular limiting magnitude.)
}
\label{fig:completeness}
\end{figure}

\begin{figure}
\centering
\includegraphics[width=8.6cm]{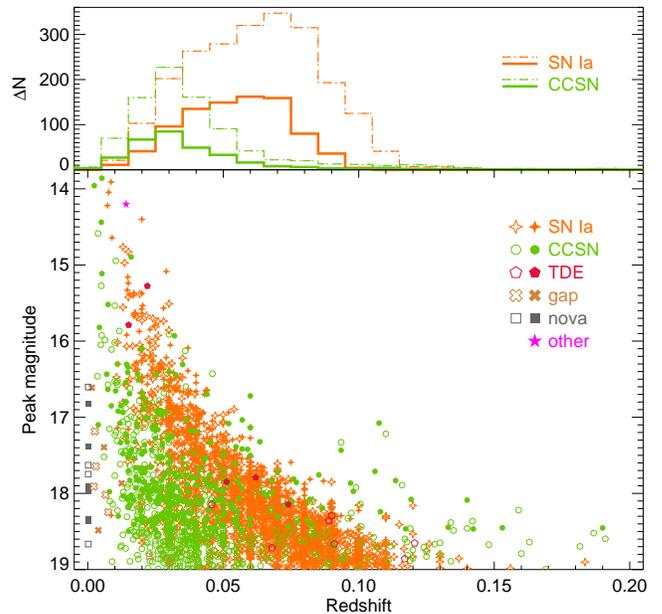}
\caption{Apparent magnitude and redshift distribution of classified transients in the BTS sample.  Symbol conventions are the same as in Figure~\ref{fig:skymap}. A histogram by redshift for Type Ia SNe and CC SNe is shown in the top panel; thick lines show $m<18.5$ mag transients satisfying all sample cuts and dashed lines indicate other classified transients.  Our survey probes CC~SNe out to approximately $z<0.05$ and SNe~Ia out to $z<0.1$, and superluminous events beyond these limits.}
\label{fig:redshiftdistribution}
\end{figure}


\begin{deluxetable}{lrrrr}
\tabletypesize{\footnotesize}
\tablecolumns{5}
\tablewidth{0pt}
\tablecaption{Classification totals \label{tab:sncount}}
\tablehead{
\colhead{} & \colhead{} & \colhead{} & \colhead{} & \colhead{Passing cuts} \vspace{-0.25cm} \\ 
\colhead{Class} & \colhead{} & \colhead{All} & \colhead{Passing cuts} & \colhead{and $m<18.5$}
}
\startdata
{\bf Transients}       & & {\bf 3147} & {\bf 1865} & {\bf 1206} \\
\phm{ss}SN Ia          & & 2232 & 1352 &  875 \\
\phm{ss}SN CC	       & &  878 &  490 &  313 \\
\phm{sssss}    H-rich  & &  671 &  357 &  226 \\
\phm{ssssssss} II      & &  516 &  273 &  171 \\
\phm{ssssssss} IIb     & &   45 &   28 &   15 \\
\phm{ssssssss} IIn     & &   89 &   45 &   32 \\
\phm{ssssssss} SLSN-II & &   21 &   11 &    8 \\
\phm{sssss}    H-poor  & &  207 &  133 &   87 \\ 
\phm{ssssssss} Ib/Ic   & &  141 &   86 &   51 \\
\phm{ssssssss} Ic-BL   & &   27 &   21 &   17 \\
\phm{ssssssss} Ibn     & &   11 &    9 &    8 \\
\phm{ssssssss} SLSN-I  & &   28 &   17 &   11 \\
\phm{ss}TDE	           & &   13 &    8 &    5 \\
\phm{ss}Gap	           & &   11 &    5 &    4 \\
\phm{ss}Novae	       & &   11 &    8 &    7 \\
\phm{ss}Other   	   & &    2 &    2 &    2 \\
Unclassified           & & 1596 &  627 &   82 \\
\enddata
\tablecomments{Totals include only public classifications available on TNS or other open sources and should be considered preliminary, pending reanalysis.  ``Gap'' transients include Ca-rich events, luminous red novae, intermediate-luminosity red transients, and LBV eruptions.  ``Other'' transients
are AT2018cow and AT2019cmw, which have extensive spectroscopy but resemble no well-established transient type and likely belong to new categories of object.}
\end{deluxetable}

\begin{figure}
\centering
\includegraphics[width=8.6cm]{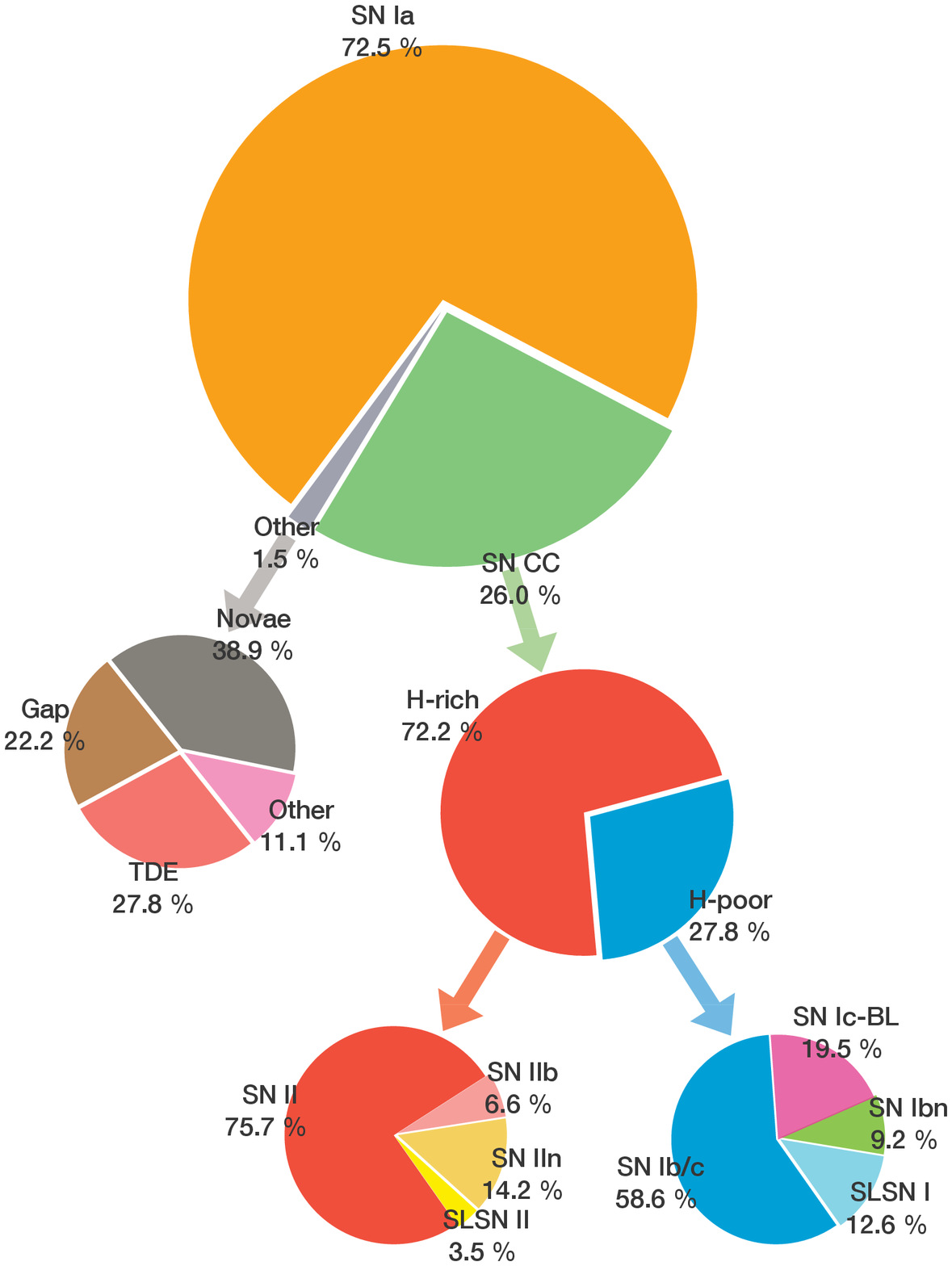}
\caption{Pie charts showing fractional number counts in the $m<18.5$ mag BTS statistical sample (1206 events in total) within various categories and subcategories.}
\label{fig:piechart}
\end{figure}

To check this we performed several tests.  First, we took every event saved to the program satisfying the quality and purity cuts that is not classified as a star/AGN (or other false positive); we refer to this population as the ``statistical sample.''  A histogram of these objects by peak magnitude is shown in Figure~\ref{fig:completeness}.  The overall distribution of these events (solid black line) is consistent with the $\Delta N \propto f^{3/2}$ power law predicted for a flux-limited survey in a homogeneous, Euclidean universe, except for the faintest bin ($m>18.9$ mag) where we anticipate being incomplete. This indicates that we are saving the expected numbers of events independent of their magnitude.

To quantify our success at spectroscopically classifying this population, a histogram of classified events is also shown in Figure~\ref{fig:completeness}.  Expressed in cumulative terms (dotted line in the upper panel of Figure~\ref{fig:completeness}), 93\% of events passing our selection cuts at $m<18.5$ mag have successful, public classifications.
Completeness improves to 97\% at $m<18$ mag and 100\% at $m<17$ mag.  Since we do not systematically target events fainter than $m>18.5$ mag, classification completeness drops sharply beyond this point: to 85\% at $m<18.75$ mag and 75\% at $m<19$ mag.

Inspection of the light curves and locations of unclassified events (at $m<18.5$ mag) confirms that most are likely to be ordinary SNe with properties that generally reflect the demographics of the rest of the sample (i.e., most appear to be SNe~Ia) and were missed solely because no spectrum could be obtained or was obtained and had low quality, usually in association with periods of bad weather (Appendix~\ref{sec:season}).  A few unclassified events have peculiar light curves that do not resemble any known SN type, including a handful with only single-night detections which may be particularly fast transients (but could also be the result of data-quality issues affecting the rest of the light curve).  A discussion of these will be deferred until forced photometry is available at the end of the survey.

To check for events we may have failed to save to our program in the first place, we downloaded the entire catalog of classified TNS transients reported between 2018-01-01 and 2020-08-12 with declination $\delta > -30 \deg$ (4350 in total).  Of these, 1015 were not in the BTS catalog.  We cross-matched the coordinates against the complete database of public-program ZTF Avro alerts using the GROWTH Marshal and, for all matches, checked how many had exceeded 19 mag in ZTF public data in more than two observations.  After removing matches that were not actually SNe we found 75 missing transients.  Most of these either (a) passed the filter on only a single night owing to extremely sparse coverage and/or a light-curve peak just above $m=19$ mag, (b) occurred during the spring 2018 science validation period before the formal public survey began, (c) were contaminated by SN light in the reference image or by a coincident foreground star, or (d) were close to bright foreground stars.  These would, by design, not have passed our alert filter and/or our selection cuts.  We did identify 13 SNe which nominally satisfy our selection goals (eight at $m<18.5$ mag and five at $18.5<m<19$ mag).  Five passed the alert filter but were not saved by scanners. 
The remaining eight did not pass the filter because the nuclei of their host galaxies were incorrectly treated as stars owing to a high or ambiguous \texttt{sgscore}.
These represent possible examples of incompleteness (with those in the latter category also potentially imposing some bias, since they are close to bright galaxy nuclei).  However, they represent a very small fraction of the overall sample total (0.6\% of $m_{\rm pk}<18.5$ mag statistical-sample transients).  Even accounting for the likelihood that some additional bright SNe may have been missed by both us \emph{and} the broader community, we expect any impact on our scientific goals to be minimal.

The above checks suggest that our program has largely succeeded in reaching its goal of producing a magnitude-limited sample with no significant biases relating to the duration, behavior, or host-galaxy environment of the transient.  

The statistics above are appropriate for ``ordinary'' SNe of the type that dominate the overall transient rate.  We do expect to fare less well in other circumstances.  Fast events which mimic CVs (and are either not in galaxies or are in galaxy nuclei), SNe coincident ($\lesssim 1$ arcsec) with AGNs, or long-timescale transients from the central regions of galaxies which resemble AGNs are all unlikely to be classified by other observers for the same reasons that they are much more likely to be missed by our selection process or excluded by our sample cuts.   
Certain rare transient categories are particularly likely to be heavily impacted:  fast and luminous transients at high redshift such as on-axis GRB afterglows, or transients specific to galaxy nuclei such as tidal disruption events (TDEs).  Finding these events effectively and studying their demographics in an unbiased way requires different selection methods.  Parallel efforts within ZTF focusing on these populations are ongoing; these are described in other works \citep{Ho+2020blt,vanVelzen2019,vanVelzen2020}.

Additionally, as previously noted, the limited cadence of the ZTF public survey itself results in a milder bias against the shortest-duration transients ($<$\,10 days, and especially $<$\,3 days) and longest-duration transients ($>$\,200 days).  Although we are still sensitive to transients with these properties, additional corrections would be necessary to accurately calculate their rates or study their demographics in a complete sense.

The redshift and magnitude distribution of the sample, before and after applying the selection cuts, is presented in Figure~\ref{fig:redshiftdistribution}.  A breakdown of transients by classification category is presented in Table~\ref{tab:sncount} and Figure~\ref{fig:piechart}.

\section{Results}
\label{sec:results}

The BTS catalog is the first large, highly complete, untargeted sample of transients for which spectroscopy and high-quality light curves are simultaneously available.   This provides opportunities to examine the complete observational parameter space occupied by these explosions, and to examine correlations between key parameters of interest, without being limited by selection bias.

A complete investigation of all potential scientific uses of this sample is beyond the immediate scope of this paper, and will be reserved for a variety of follow-up works once the first phase of ZTF is finished and complete forced-photometry light curves are available, along with final spectroscopic (re-)classifications.  {A complete and impartial spectroscopic analysis will be essential for uncommon or easily-confused subclasses in particular (e.g., Ia-CSM vs. IIn, II vs. IIb, or Ib vs. Ic)}.  For the vast majority of transients, however, neither the light curves nor the classifications are expected to change significantly.   In this section we will make use of the existing data products to provide a preliminary exploration of a variety of topics to demonstrate the scientific capabilities of the sample.

\subsection{The Landscape of Stellar Death}
\label{sec:landscape}

\begin{figure*}
\centering
\includegraphics[width=18cm]{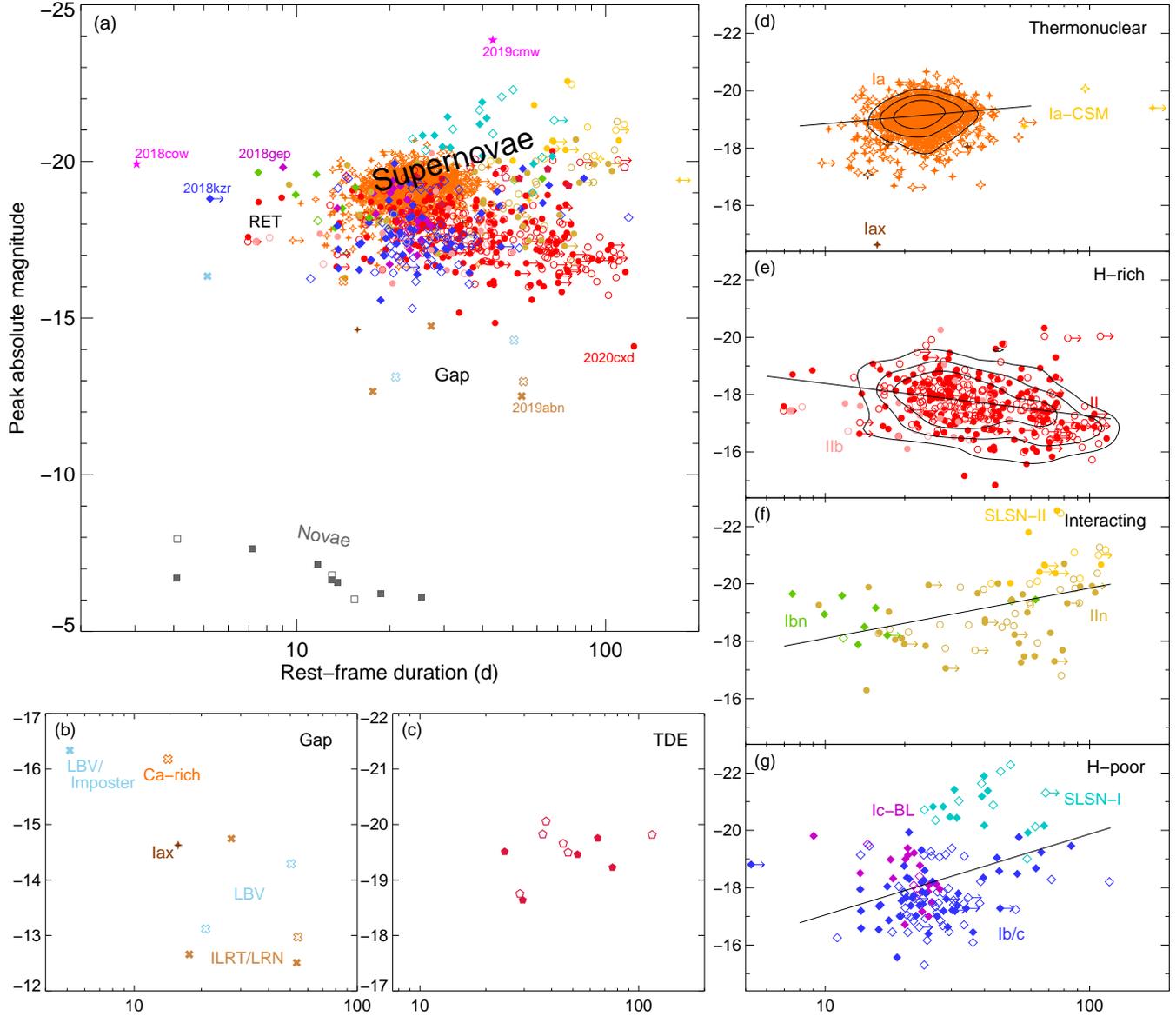}
\caption{(a) Duration-luminosity plot for 1197 classified BTS transients at $m<18.5$ mag satisfying our quality cut (filled points) and 966 additional transients with usable timescale measurements (unfilled points).  Durations are time above half-peak and absolute magnitude is at the observed peak.  The surrounding panels, (b)--(g), break the population into general spectral types, with non-SN populations on the bottom and SN populations at right.  {Panels (e)--(g) refer to CCSNe, with strongly CSM-interacting members of H-rich and H-poor populations shown in panel (f).}   Most SN types and subtypes occupy distinct (if overlapping) regions within duration-luminosity parameter space.  Contours show 50\%, 75\%, and 90\% containment of the kernel density estimate for SNe~Ia and H-rich SNe.   Correlations between duration and luminosity are observed for most SN populations.}
\label{fig:timelum}
\end{figure*}

\cite{Kasliwal2011} summarized the state of knowledge of transient ``parameter space'' (in luminosity and characteristic timescale), highlighting advances provided by wide-field surveys in discovering events with properties different from those of typical SNe:  very luminous events (SLSNe), events intermediate in luminosity between novae and typical SNe (``gap'' transients), and very fast events.  {More recently, \citet{Villar+2017} approached the issue theoretically, providing a physical explanation for the luminosity and timescale distributions of many types of known and predicted transients.}

We are now in a position to provide an unbiased look at this topic using a complete sample of real transients.  This is provided in Figure~\ref{fig:timelum}, calculated using the BTS sample.  Filled points show events that pass the quality and purity cuts and peak at $m \leq 18.5$ mag.  We also show other events as open circles as long as they either pass the quality/purity cuts \emph{or} have a useful measurement of both their rise and fade times even in the presence of poorer sampling or a fainter peak, though this supplementary sample is not unbiased and the associated measurements typically have larger uncertainties.  The timescale (rest-frame time above half-maximum light, calculated by adding the rise and fade times and dividing by the time dilation factor of $1+z$) and peak luminosity are calculated using the basic interpolation method described in \S\,\ref{sec:charac}.  Events which only have lower limits on their timescales have been omitted from the diagram if the limit is not ``constraining'' in comparison to the bulk of the SN population ($>16$ days), \emph{unless} the rise time has been measured and is $<8$ days (i.e., unless there is reason to think it is an actual fast transient whose decay was not well captured, rather than a transient with a sparse light curve).  A small number of SNe with data-quality issues identified by manual inspection were also removed.

The region of the diagram with characteristics of typical SNe (timescales of about a month and absolute magnitude close to $-18$) is extremely well populated, as expected.  However, smaller numbers of events do populate the diagram in all directions except the longest durations ($\gg 100$ days), to which we are not yet sensitive because of the limited duration of the survey.

\subsubsection{Rapidly Evolving Transients}
\label{sec:rets}

Transients in the leftmost part of the diagram ($t<10$ days) are expected to be somewhat undersampled owing to the limited cadence (\S\,\ref{sec:completeness}).   Even so, it is clear that very fast events (sometimes called rapidly-evolving transients or RETs; \citealt{Drout+2014,Arcavi+2016,Pursiainen+2018,Wiseman+2020}\footnote{Other acronyms used include fast-evolving luminous transients (FELTs; \citealt{Rest+2018}) or fast blue optical transients (FBOTs; \citealt{Margutti+2019}).}) are quite rare: there are only fourteen with total durations of $<10$ days and peak absolute magnitudes $M<-16$ in the (statistical) sample.  The most striking such object is AT2018cow \citep{Prentice2018,Perley2019b,Ho+2018cow,Margutti+2019}, which independently passed our sample cuts and appears as an outlier in parameter space, even in comparison to the other RETs.  The physical nature of this event is still unknown.  The next-fastest luminous event in the diagram, SN~2018kzr, faded too rapidly for its timescale to be measured precisely, but its rapid nature is confirmed by its fast rise.  Nominally spectroscopically classified as a SN~Ic, this event has been suggested to be a white dwarf accretion-induced collapse or a NS-WD merger \citep{McBrien+2019,Gillanders+2020}. 

\begin{figure*}
\centering
\includegraphics[width=18cm]{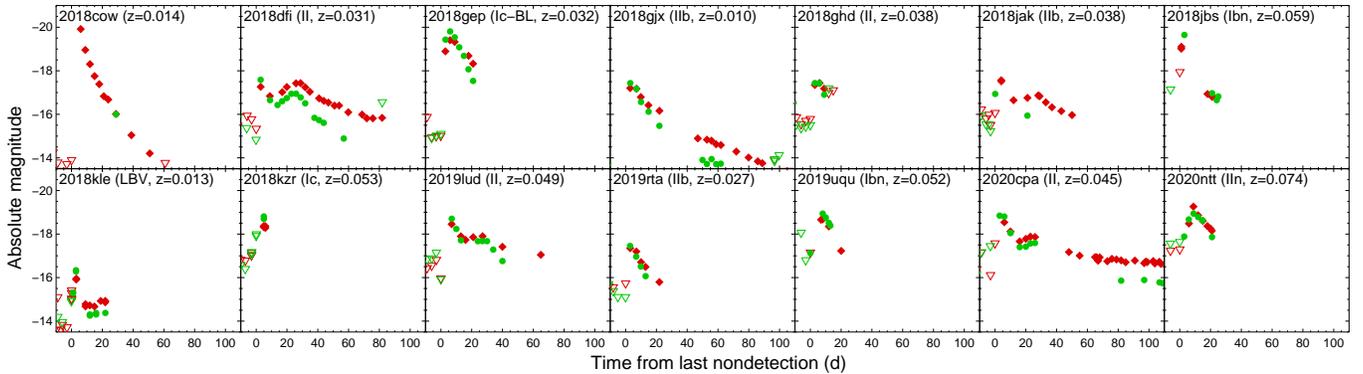}
\caption{ZTF light curves for rapidly evolving (time above half-peak $t < 10$ days) transients in the sample.  Symbol conventions are as in Figure~\ref{fig:lightcurves}.  Most fast transients belong to known SN classes and many show clear second peaks or late-time slow-decline phases.  These late-time features may not be obvious in surveys operating closer to the detection limit.}
\label{fig:fastlightcurves}
\end{figure*}

The remaining twelve events (Figure~\ref{fig:fastlightcurves}) are much less extreme (all have $>5$ days duration) and show a variety of, generally ordinary, spectroscopic classifications: Type II and IIb SNe are most common (4 and 2 examples, respectively), but there are also two Type Ibn, one fast Type IIn, one peculiar Type Ic-BL (SN2018gep; \citealt{Ho+2018gep}), and one luminous blue variable (LBV) SN imposter.  Most of these events are quite luminous ($M < -18$ mag), making them incompatible with radioactive heating as the primary energy source \citep{Arnett89} for the main peak, although a few show subsequent second peaks or plateaus on a classical SN timescale.  Shock-breakout into an extended envelope or circumstellar medium (CSM) has been suggested as the likely explanation for previous fast-peaking, luminous events (both for AT2018cow and for less-extreme RETs: \citealt{Ofek+2010,Drout+2014,Pursiainen+2018,Perley2019b,Margutti+2019,Ho+2018gep}), and it is reasonable to hypothesize that this process is also responsible for most or all of the transients in this portion of the diagram in our sample.

As most of these evolved into relatively ordinary SN classes later in their evolution, it is not obvious that any additional special conditions are required beyond dense, extended CSM in most cases to produce a rapidly-evolving transient, and we infer that this is also the case for most short-timescale events in the Dark Energy Survey, Pan-STARRS, and other earlier surveys.
AT2018cow-like events, with their luminous multiwavelength emission \citep{Ho+2018cow,Ho+Koala,Margutti+2019,Coppejans+2020}, represent a dramatic but rare exception, and certain SNe may also require an engine-driven jet to power the fast, energetic shock inferred from radio and X-ray observations \citep{Ho+2020bvc}.

\subsubsection{Low-Luminosity (``Gap'') Transients}
\label{sec:gap}

The low-luminosity region of the diagram in the ``gap'' between novae and SNe ($-9 > M > -15$ mag) is sparsely sampled --- as expected for a magnitude-limited survey, since the volume to which events with these properties can be detected is very limited.   We do in fact detect similar numbers of low-luminosity transients and classical SNe at very low redshifts ({two of each at} $z<0.004$ or $d<17$ Mpc, where we are complete to $M_{\rm lim} \approx -12.7$ mag), suggesting that the volumetric rates of dim transients are at least comparable to those of classical SNe (see also \citealt{Frohmaier+2018}).

Some low-luminosity events are clearly SNe themselves. SN2020cxd at $M_{\rm peak}=-14.1$ mag is the most notable such example; with H$\alpha$ velocity widths of 5000 km s$^{-1}$ in its spectrum and a SN~IIP-like light curve, it is almost certainly the explosion of a massive star, despite being an outlier relative to other Type II SNe in the sample.  A handful of similar events are known in the literature \citep[e.g.,][]{Pastorello+2004,Pastorello+2009,GalYam+2011}. 

The remaining objects in the ``gap'' luminosity band do not match well-established SN templates.  The classification system for these types of events is still evolving, and the progenitor interpretation of these classes remains an active area of research.  Two are LBVs in a very luminous, high-activity state\footnote{This category stretches our definition of ``transient,'' and our catalog is unlikely to be complete to such events.}. One is classified as an SN 2002cx-like SN Ia (``SN Iax''; \citealt{Li+2003,Foley+2009,Foley+2013}), possibly an incomplete SN~Ia that does not fully destroy the white dwarf.    One is classified as a luminous red nova (LRN), a class of event generally interpreted as stellar mergers \citep{Kulkarni+2007,Pejcha+2016,Pastorello+2019lrn,Blagorodnova+2020}.  Three are classified as intermediate-luminosity red transients (ILRTs), a broadly-defined observational class sometimes attributed to electron-capture SNe \citep{Thompson+2009,Botticella+2009,Moriya2014}, although this remains controversial; the distinction between these events and LRNs is not always obvious \citep{Cai+2019}, and a variety of other models exist \citep{Pastorello+2007,Bond+2009,Berger+2009,Tsuna+2020}.  The lowest-luminosity non-nova transient in the BTS sample, ILRT AT2019abn, is discussed in detail by \cite{Jencson+2019}.

A few additional events in our sample belong spectroscopically to low-luminosity classes that are outside the traditional SN scheme, but which individually have significantly higher luminosities and overlap with the SN distribution ($M<-15$ mag).  Among Ca-rich events \citep{Filippenko03,Perets+2010,Kasliwal12} in BTS, only SN 2019hty 
\citep{De2020CLU}, with $M=-16.1$ mag, passed our selection cuts.  There is also one particularly luminous LBV eruption (SN imposter; \citealt{Maund2006,Smith+2011,Pastorello+2019gap}), and several luminous SNe~Iax.   A compilation of all events that are not standard SN types and have luminosities between $-12 < M <-17$ mag are shown in Figure~\ref{fig:timelum}b, with SNe~Iax shown in comparison to the general SN~Ia distribution in Figure~\ref{fig:timelum}d.  

At this stage the sample of low-luminosity events remains too small for a detailed examination of their population properties.  The ZTF volume-limited survey (CLU) provides a much larger sample of transients in this regime,  and a significantly expanded discussion of this population will be provided in forthcoming work in association with that effort.  An analysis of the hydrogen-poor subset of low-luminosity transients from the first two years of ZTF, with an emphasis on Ca-rich events, can be found in \citet{De2020CLU}.  A discussion of low-luminosity, hydrogen-rich SNe will be provided by Tzanidakis et al. (in prep.).

\subsubsection{Superluminous Transients}
\label{sec:slsn}

At the high-luminosity end, the superluminous supernova (SLSN; for reviews see \citealt{Gal-Yam12,GalYam2019}) population is clearly visible as a group extending to the top and right of the general SN population, although with no indication of a gap between the SLSN and general SN populations (in agreement with \citealt{DeCia+2018}).  Hydrogen-poor and hydrogen-rich SLSNe form distinct regions in duration space:  SLSNe-II are universally longer in duration.  The SLSN population has been the focus of intensive efforts within ZTF to provide high completeness to significantly deeper limits than $m>18.5$ mag (e.g., \citealt{Lunnan2019}), and further discussion of this population will be reserved for a series of upcoming papers by Yan et al., Perley et al., and Chen et al.

No securely-classified transient in the BTS sample is more luminous than $M<-23$ mag, although we have found one featureless, slow transient with $M=-23.6$ mag (AT2019cmw) inferred from a redshift measurement via intergalactic-medium absorption lines.  It is not yet clear whether this represents an extreme SN, a TDE, or a particularly extreme AGN accretion phenomenon; it will be addressed by a subsequent study.

\subsubsection{Tidal Disruption Events}

The number of TDEs within the BTS sample is relatively small (Figure~\ref{fig:timelum}c) and as of yet insufficient for a detailed statistical investigation.   While all are quite luminous ($M_{\rm peak} < -18$ mag, and all but two are at $M_{\rm peak} < -19$ mag), the absence of lower-luminosity examples does not yet rule out the possibility that fainter events \citep[e.g.,][]{Blagorodnova+2017} comprise the bulk of the population.   Timescales range between approximately 20--120 days, although we are unlikely to be sensitive to any longer-duration events as they would be indistinguishable from AGNs to our filter (\S\,\ref{sec:filter}).   A more complete overview of TDEs within ZTF can be found in the sample study of \cite{vanVelzen2020}.

\subsection{SNe and Luminosity-Duration Correlations}
\label{sec:sne}

Events traditionally defined as SNe broadly occupy a common region of the diagram (of typically 10--100 days duration and absolute magnitudes between $-16$ and $-21$) and there is overlap between all SN types in this region.  Even so, there are clear distinctions between different classes in this parameter space, with different trends emerging among different groups.  In the right four panels [(d)--(g)] of Figure~\ref{fig:timelum} we have separated SNe into four general categories: thermonuclear (Ia), H-rich (``ordinary'' II and IIb), interacting (IIn, SLSN-II, and Ibn), and H-poor (Ib/c and SLSN-I); see, e.g., \citet{Filippenko1997} for a review of SN spectral classification.

Type Ia SNe, as expected, have relatively standard properties and cluster in a small locus (although additional scatter is introduced due to host-galaxy extinction, distance uncertainties, sampling gaps, and the use of both $g$ and $r$ data to determine timescales).  While our general duration parameterization differs from the $\Delta m_{15}$ decline parameter often used in SN cosmology, the well-known \cite{Phillips1993} correlation between timescale and luminosity is nevertheless qualitatively replicated at high statistical significance: a simple linear regression to the SN~Ia population shown in Figure~\ref{fig:timelum} gives a slope ($b \equiv \Delta m/\Delta {\rm log}\,t$) of $-0.80 \pm 0.15$ mag dex$^{-1}$ ($1\sigma$ uncertainties).

Interestingly, other transient classes also display similar correlations.  In particular, Type II SNe show a reasonably tight (Pearson coefficient $r=0.32$) correlation between magnitude and luminosity: longer events are dimmer ($b = 1.14 \pm 0.21$).  This is in agreement with other studies based on much smaller samples \citep{Anderson2014,Sanders+2015,Rubin+2016,Valenti+2016,Galbany+2016a,deJaeger+2019}.

Type IIn SNe (which for this purpose we take to include all events classified as SLSN-II on TNS, most but not all of which show SN~IIn-like narrow features) obey the opposite correlation: slower-declining events are on average \emph{more} luminous, in agreement with \cite{Ofek+2014} and \cite{Nyholm2020}.  This relation may turn over toward the short-timescale end of the diagram: Type Ibn SNe form a small cluster of luminous CSM-interacting hydrogen-poor transients of which the shortest events are generally \emph{more} luminous than the longer events, but the sample is small and the trend is not significant.   A linear fit to the entire interacting SN population as shown in Figure~\ref{fig:timelum} gives $b = -1.75 \pm 0.42$ and $r = -0.50$.

Hydrogen-poor SNe, like interacting SNe, show a positive correlation between duration and luminosity ($b = -2.82 \pm 0.64$, $r=-0.34$).  There is a hint that this population may cluster into separate subpopulations:  a cluster of ordinary SNe~Ib/c but also a population of much slower and brighter SNe~Ic including SLSNe-I.  A $k$-means clustering analysis did not confirm that these clusters are statistically significant, so larger samples will be necessary to confirm this hypothesis.

These correlations may have useful cosmological applications, and the distinguishability of different subpopulations (even without color or shape information) may be encouraging for the use of photometric techniques to classify transients in future surveys.  They may also provide further insight into the physics and progenitor populations of explosions of different types.

\subsection{Rate Measurements}
\label{sec:rates}

\begin{figure*}
\centering
\includegraphics[width=18cm]{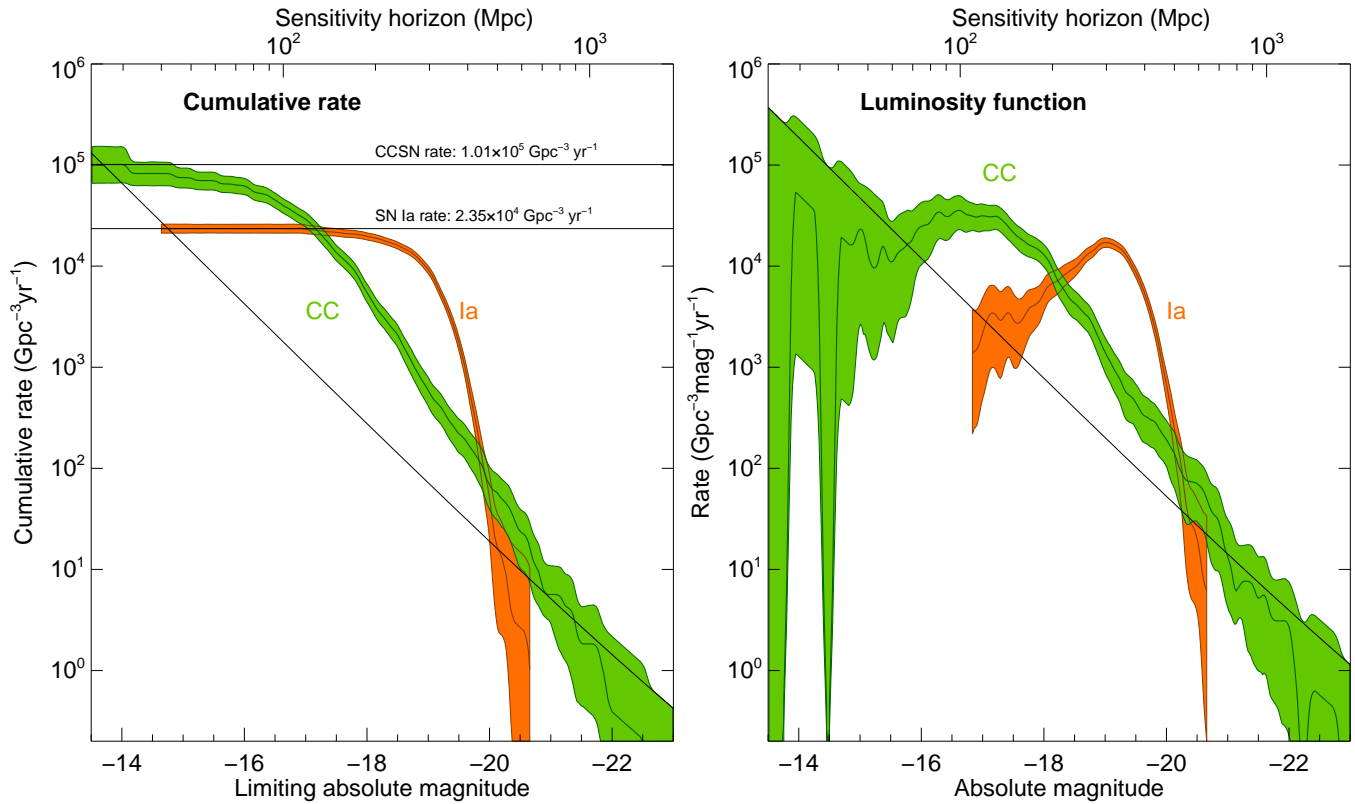}
\caption{Rate measurements for Type Ia and CC~SNe.  The plot on the left shows the volumetric rate of SNe brighter than a particular absolute magnitude (in either the $g$ or $r$ band, without correcting for host-galaxy extinction); the plot on the right shows the luminosity function calculated from SNe peaking within $\pm0.25$ mag of a particular magnitude.  {Colored bands correspond to 95\% statistical confidence intervals.}  Horizontal lines show the total SN~Ia rate (to $-16.5$ mag) and CC~SN rate (to $-14$ mag) estimated from this work.  The black diagonal line indicates statistical upper limits (95\% confidence) for the case of zero detected events at a given magnitude.}
\label{fig:rates}
\end{figure*}

Another key advantage of a magnitude-limited, untargeted, spectroscopically-complete survey is that the measurement of volumetric rates is relatively straightforward.  

For an ideal survey that is able to scan the entire sky to a given magnitude limit $m_{\rm lim}$ without interruption and  unaffected by Galactic dust, collecting a sample of $N$ events over a total survey time $T$ much longer than the duration of any individual transient, the volumetric rate can be estimated from

\begin{equation}
    R = \frac{1}{t} \sum_{i=1}^{N} \frac{1}{(\frac{4\pi}{3}D_{{\rm max},i}^3)},
\end{equation}

\noindent
where $D_{{\rm max},i}$ is the distance out to which the $i$th transient can be detected above $m_{\rm lim}$ at peak light in the absence of extinction, given its peak absolute magnitude $M_i$.

Any real SN survey does not cover the whole sky, operates over a finite time window with a complex cadence structure, must contend with Galactic extinction, and does not recover all of the transients it ``detects.''  This requires additional correction terms to compensate for the effective loss of survey volume, and for the gain of additional transients that ``occurred'' (peaked) outside the survey time window but were detected on the rise or the decline.  These corrections can potentially be difficult to apply in practice since the loss/gain factors may vary by transient type, sky location, and other factors.  In the case of BTS, we have strictly chosen a sample such that the peak is well determined using unbiased sample cuts and guaranteed to occur within the survey window, making this task much simpler.  A revised equation is

\begin{equation}
    R = \frac{1}{T} \sum_{i=1}^{N} \frac{1}{(\frac{4\pi}{3}D_{{\rm max},i}^3)  f_{\rm sky} f_{\rm ext} f_{\rm rec}  f_{{\rm cl},i}}.
\end{equation}
\noindent
The loss factors are as follows.

\begin{itemize}
\item $f_{\rm sky}$ is the average active survey footprint expressed as a fraction of the full sky. 
\item $f_{\rm ext}$ is the average reduction in effective survey volume owing to Galactic extinction.
\item $f_{\rm rec}$ is the average recovery efficiency for a detectable transient within the survey footprint: the probability that it is found and included in the sample.  
\item $f_{{\rm cl},i}$ is the classification efficiency.  (This may depend on apparent magnitude, so the subscript is retained.)
\end{itemize}

Using the exposure history from the public survey, we estimate the average active area across the three-night cadence cycle over the period considered here of 14400 deg$^2$ ($f_{\rm sky} = 0.35$).

The Galactic extinction correction $f_{\rm ext}$ can be calculated by averaging the reduction in volume associated with the extinction toward each separate ZTF field (excluding fields with $A_V<1.0$ mag that are omitted from the sample).  We infer $f_{\rm ext}=0.82$.

The recovery fraction is the most uncertain parameter.  
We previously estimated (\S\,\ref{sec:quality}) that 52\% of our candidate transients passed our quality cuts, but this is not an ideal estimate because some candidates were not transients, or may have occurred (peaked) outside the active sky region and been classified much later.  To provide a better estimate of this parameter, we took all events classified as SNe~Ia with peak absolute magnitudes of $<-18.5$ and peak apparent magnitudes of $<18$, a set of conditions that effectively ensures that the transient peaked within the active area and that it would have been very easy to classify even in suboptimal observing circumstances.  Of these, 412/690 pass the quality cut, so we estimate $f_{\rm rec}=0.60$.

Our classification completeness was addressed in \S\,\ref{sec:completeness}.  It is close to 100\% for bright transients but declines to about 90\% at $m=18.5$ mag and drops quickly afterward.  We assume $f_{\rm cl}$=1.0 if $m<17.2$ mag and $f_{\rm cl}$=0.9 at $m=18.5$ mag, with a linear decline in between.

The sample as presented in this paper spans $t = 2.12$ yr of ZTF.  We assume a uniform $K$-correction ($K = 2.5\times {\rm log}_{10}(1+z)$) and ignore cosmological effects.\footnote{The contraction of the control time window ($\Delta t \propto(1+z)^{-1}$) is approximately compensated for by the increase of the star-formation rate density ({\rm SFR}$\propto(1+z)^{+1.2}$ in the low-redshift limit of Equation 9 of \citealt{Madau+2014}) and redshift-dependent SN rates \citep[e.g.,][]{Dahlen+2004,Dahlen+2008,Barris+2006,Melinder+2012,Strolger+2015,Graur+2011,Dilday+2010}.}
 
Based on these assumptions (and $H_0=70$ km s$^{-1}$ Mpc$^{-1}$), we infer a SN~Ia rate of $(2.35 \pm 0.24)\times10^{4}$ Gpc$^{-3}$ yr$^{-1}$ (95\% confidence interval, statistical errors only).  Caution should be taken in interpreting this as a truly independent measurement of the SN~Ia rate: the survey parameters above were not chosen entirely blindly of the result and the true uncertainty will be dominated by systematics, which are not easy to quantify.  Even so, it is encouraging that this value is very close to the value from several large-scale studies over the past ten years \citep{Dilday+2010,Li11,Graur+2011,Frohmaier2019}.  We are not able to confirm the claim by \cite{Smith+2019} of a much higher rate.

This method can be generalized to any population and any limiting magnitude or magnitude range; in Figure~\ref{fig:rates} we show both cumulative rates and luminosity functions for the SN~Ia and CC~SN populations.  We replace the lower limit with a limit calculated from the $m<19.0$ mag sample (a limiting value of $f_{\rm cl}=0.9$ is assumed beyond $m>18.5$ mag) if this is more constraining than that from the $m<18.5$ mag sample.  Note that luminosities are as observed: host-galaxy extinction is \emph{not} removed, although Galactic extinction is corrected for. 

Calculating the total CC~SN rate is more challenging than calculating the SN~Ia rate, both because the number counts are less but also because the luminosity function is broader, with a significant fraction of the population coming from very dim events that are not detectable except in small volumes within the nearby universe (see also \citealt{Taylor+2014} and forthcoming work by Tzanidakis et al.).  Assuming a minimum luminosity of $M < -14$ mag we infer a rate of $(10.1^{+5.0}_{-3.5})\times10^{4}$ Gpc$^{-3}$ yr$^{-1}$.

This value is fully consistent with predictions based on the low-redshift star-formation rate density (\cite{Madau+2014}).  This is in agreement with other works arguing that the ``SN rate problem'' (\citealt{Horiuchi+2011}, originally motivated by the lower CC~SN rate of \citealt{Li11}) is resolved using galaxy-untargeted surveys and including the faint end of the luminosity function, without requiring a large population of completely optically-obscured SNe (although such events may exist; \citealt{Mattila+2012,Jencson+2019}) or direct collapses.

This same methodology could be applied to other, rarer transients in the sample.  For example, inspection of the high-luminosity end of the CC~SN rate curve implies an SLSN rate (above $M<-21$ mag) of $5.6^{+5.4}_{-2.8}$ Gpc$^{-3}$ yr$^{-1}$, which is much lower than the commonly cited estimate by \cite{Quimby+2013} of $199^{+137}_{-86}$ Gpc$^{-3}$ yr$^{-1}$)\footnote{It is also lower than the estimate at $z=1.1$ by \citet{Prajs+2017} of $91^{+76}_{-36}$ Gpc$^{-3}$ yr$^{-1}$, although if the factor of $\sim6$ increase in the star-formation rate density with cosmic time is taken into account, our estimates are marginally consistent within the uncertainties.}.  We will present further calculations of the SLSN rate and luminosity function (using the full ZTF SLSN sample of $>150$ events, extending to much fainter limiting magnitudes and with appropriate cosmological and $K$-corrections) in forthcoming work by Yan et al.

Rate calculations can, in principle, be extended even to classes of transients we do not detect at all.  The diagonal lines in Figure~\ref{fig:rates} show 95\% confidence upper limits on the intrinsic rate of any transient for which we have found no examples in BTS so far, assuming that the transient does not have properties that make it systematically selected against by the survey cadence or our selection cuts (e.g., very short duration, occurs near variable AGNs, etc.) and that it is not mistaken for another class of object.  For example, we have not detected any event with properties consistent with an off-axis GRB afterglow or a kilonova in BTS to date.  Assuming that these events are not selectively missed, this would imply that the rate and luminosity functions lie below these limits; e.g., $<5\times10^3$ Gpc$^{-3}$ yr$^{-1}$ for kilonovae more luminous than $M<-16$ mag, or $<7$ Gpc$^{-3}$ yr$^{-1}$ for off-axis afterglows peaking above $M<-20.5$ mag.  The kilonova estimate is consistent with the rate estimated by \cite{Andreoni2020} using all ZTF data although not as constraining, since \cite{Andreoni2020} do not require spectroscopic classifications and probe much deeper than $m>18.5$ mag.  The GRB rate is consistent with expectations given the on-axis GRB rate of $\sim1$ Gpc$^{-3}$ yr$^{-1}$ \citep{Wanderman+2010}, assuming we would only be sensitive to events seen within a viewing angle of $\theta < 3 \theta_{\rm jet}$ at this luminosity threshold.  

\subsection{Host-Galaxy Properties}
\label{sec:hosts}

\begin{figure*}
    \begin{minipage}{0.49\textwidth}
        \centering
        \includegraphics[width=0.99\textwidth]{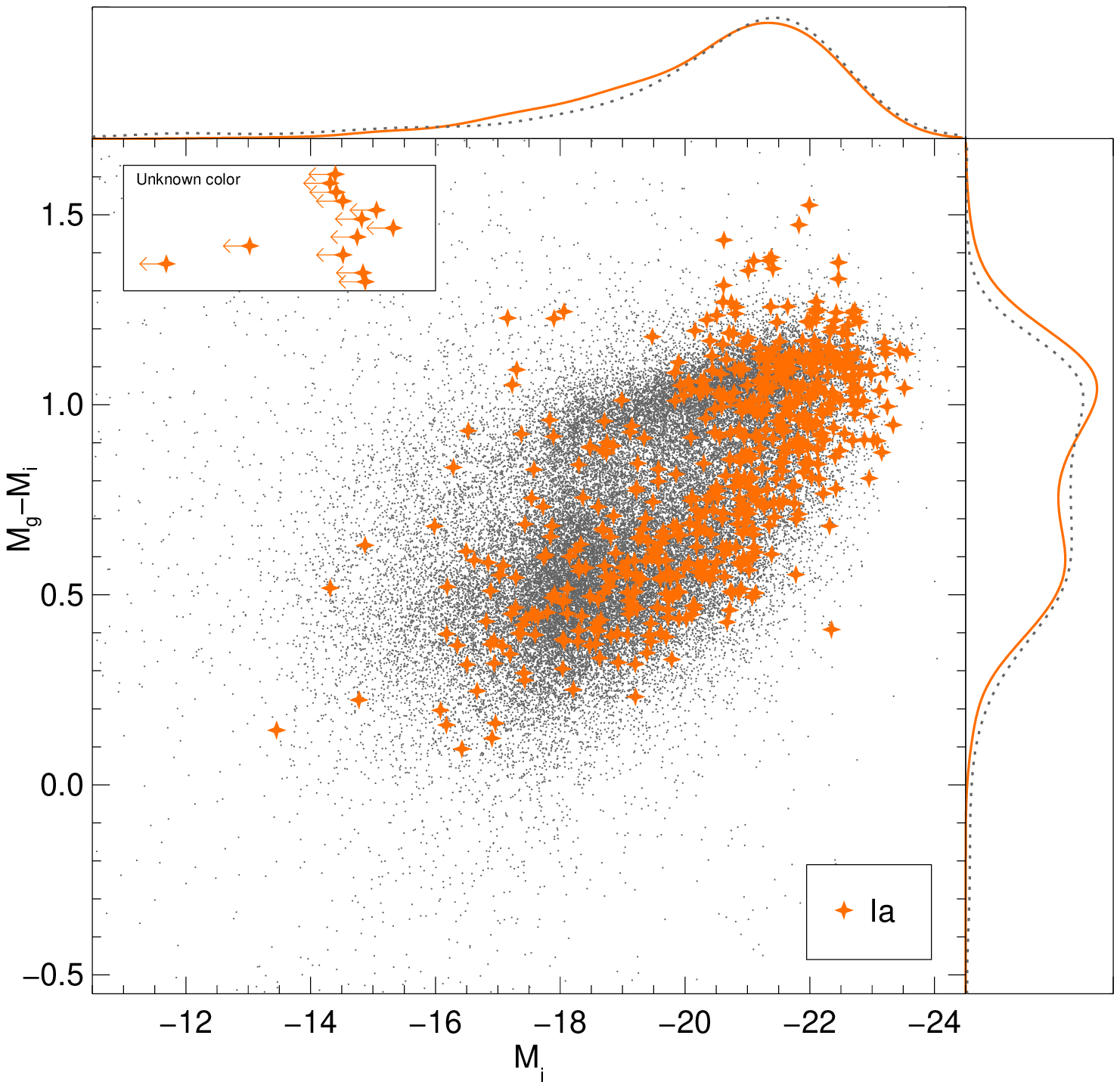}
    \end{minipage}\hfill
    \begin{minipage}{0.49\textwidth}
        \centering
        \includegraphics[width=0.99\textwidth]{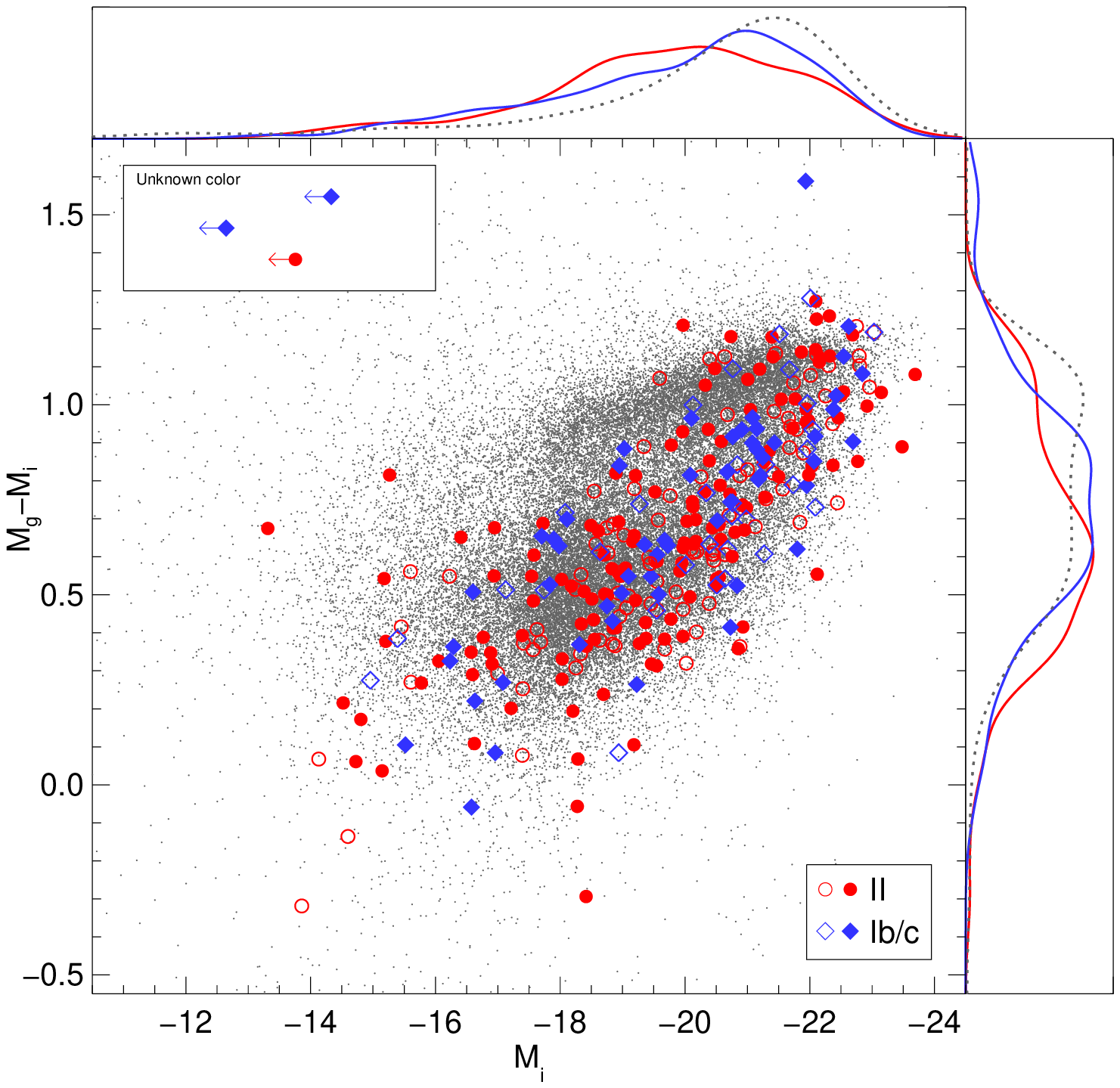}
    \end{minipage}
    \caption{Color-magnitude diagrams (rest-frame $g-i$ color versus $r$-band absolute magnitude) for the host galaxies of SNe~Ia (left) and of CC~SNe (right).  We restrict the SN host population to a redshift range of $0.015<z<0.1$ and show only events in SDSS fields passing the quality cuts.  For SNe~Ia we show only $m<18.5$ mag events but for CCSNe we also show host galaxies of fainter events (unfilled symbols).  SDSS galaxies from the NASA-SDSS Atlas at $z<0.03$ are shown as gray points.  The inset box at upper left shows SNe with no host association, which have a magnitude upper limit but no color constraint (their $y$-axis positions within the inset are arbitrary).  The side panels show kernel-density curves for each SN host population and for the SDSS galaxies (weighted by $u$-band luminosity divided by $V_{\rm max}$).   CC~SNe are grouped into only two general classes: SNe~II (including all subtypes) in red and SNe~Ib/c (including all subtypes) in blue.  The luminosity and color differences between the Type II and Type Ib/c host-galaxy populations are not statistially significant.}
    \label{fig:hostgalaxies}
\end{figure*}

The host-galaxy population of a specific transient subtype offers a valuable clue into the nature of its progenitor.  The hosts of SNe~Ia provide a means to estimate the distribution of the delay time between formation and explosion, and thereby the nature of the progenitor binary system.  The association of CC~SNe with generally young stellar populations is well established, as expected for massive stars, but measuring the (much shorter) delay time sufficiently precisely to constrain theory is much more challenging when only galaxy-integrated measurements are available.
However, comparisons between different CC~SN subtypes can still provide a powerful constraint on their respective origins.  In particular, single-star evolutionary models generally predict that mass loss will be more effective at high metallicity than at low metallicity \citep{Maeder+2000}, increasing the fraction of stripped-envelope SNe in metal-rich galaxies.  Very low metallicities have been suggested to be conducive to powering engine-powered and other rare transients that require a hydrogen-poor progenitor \emph{without} strong winds \citep[e.g.,][]{Yoon+2005}.

The BTS volume extends well beyond the redshifts at which spectroscopic galaxy catalogs are complete \citep{Fremling2020}, and a thorough statistical investigation of the transient--host-galaxy connection will require significant spectroscopic follow-up observations and spectral energy distribution (SED) fitting to aperture-matched multiwavelength data.  However, BTS redshifts are well within the range at which all-sky \emph{photometric} galaxy catalogs are largely complete for all but the lowest-luminosity galaxies, making it possible to study the basic characteristics of the sample with these data alone.

In Figure~\ref{fig:hostgalaxies} we plot the luminosities ($M_i$) and colors ($g-i$) of the host galaxies for 593 SNe~Ia and 321 CC~SNe at $0.015<z<0.1$ within the SDSS footprint\footnote{PS1 catalog photometry is affected by aperture differences and galaxy shredding to a much greater extent than for SDSS and was found to be unreliable for this purpose, so we restrict our analysis to SDSS fields for now.  We introduce a lower limit on the redshift to avoid background oversubtraction and host-galaxy mismatches, and an upper limit to ensure that nondetections are always constraining.}, calculated using a probabilistic cross-match (and subsequent manual vetting).  Both values are corrected for Galactic extinction.  For comparison, we also plot the full set of SDSS (spectroscopic) galaxies at $z<0.03$ from the NASA-SDSS Atlas \citep{Blanton+2011}.  The completeness limit for the SDSS spectroscopic sample is approximately $M_{i} < -17.9$ within this volume.

Low-redshift field-selected galaxies show a bimodal color-magnitude distribution: a ``red sequence'' dominated by passive, early-type galaxies and a ``blue cloud'' dominated by spirals and irregulars.  (The region between these populations is sometimes termed the ``green valley.'')  This is evident in the SDSS field population in Figure~\ref{fig:hostgalaxies}.  It is less obvious in the SN hosts, although the SN~Ia color distribution (right subpanel of left panel) shows the associated bimodality clearly.

\begin{deluxetable}{llrcrc}
\tabletypesize{\footnotesize}
\tablecolumns{6}
\tablewidth{0pt}
\tablecaption{Contributions of different galaxy populations to SNe~Ia and CC~SNe for SDSS fields.}
\tablehead{
\colhead{Population} & & \colhead{$N_{Ia}$} &\colhead{$f_{Ia}$} & \colhead{$N_{CC}$} & \colhead{$f_{CC}$}}
\startdata
Red sequence$^{\rm a}$ &  & 181 & 31$\pm$2 \%           & 21  & 11$\pm$2 \% \\
Green valley$^{\rm b}$ &  & 112 & 19$\pm$2 \%           & 37  & 19$\pm$3 \%\\
Blue cloud$^{\rm c}$   &  & 281 & 47$\pm$2 \%           & 128 & 64$\pm$3 \%	 \\
Subdwarf$^{\rm d}$     &  &   6 & 1.0$\pm$0.4 \%        & 10  & 5.0$\pm$1.6	\% \\
Hostless$^{\rm e}$     &  &  13 & 2.2$\pm$0.6 \%        & 3   & 1.5$^{+1.1}_{-0.7}$	\% \\
\enddata
\tablecomments{\\
$^{\rm a}$ $M_i < -16$ and $g-i > \rm{max}\{0.85-0.05(M_i+18),0.85\}$ \\ 
$^{\rm b}$ $M_i < -16$ and $0.85 < g-i < 0.85-0.05(M_i+18)$  \\
$^{\rm c}$ $M_i < -16$ and $g-i < 0.85$  \\
$^{\rm d}$ $M_i > -16$ \\
$^{\rm e}$ No host association found.  May include SNe at large offset and intracluster SNe.\\
Note --- Contributions are given both as total counts and as a fraction.  Uncertainties are approximate 1$\sigma$ binomial confidence intervals.
}
\label{tab:hostcounts}
\end{deluxetable}

We subdivided the SN host population in color-magnitude space between red-sequence, green-valley, blue-cloud, and subdwarf galaxies as defined in Table~\ref{tab:hostcounts}.  ``Hostless'' events are designated separately.  For CC~SNe these ``hostless'' events probably do have low-luminosity, coincident hosts fainter than the SDSS detection threshold (e.g., \citealt{Zinn+2012}).  For most ``hostless'' SNe~Ia we anticipate that  there is no coincident host and the progenitor has travelled a sufficient difference from the galaxy in which the system formed such that there is no probabilistically secure association, although it is possible that a few may have undetected low-luminosity, coincident hosts \citep{Strolger+2002}.

Red-sequence galaxies contribute about a third (31\%\footnote{Numbers presented in this section are based on the $m<18.5$ mag, quality-cut sample, to avoid the possibility of host-dependent spectroscopic confirmation bias.  However, we obtain consistent results for the full classified sample.}) of the SN~Ia population but also significantly contribute to the CC~SN population (11\%).   We inspected the imaging of all SDSS-matched CC~SN host galaxies with $g-i>0.9$ mag; {most have an early-type morphology or contain H~II regions or other signatures of ongoing star formation at or near the SN site, as would be expected given the short lifetimes of their progenitors.   However, two CC~SN hosts (SN 2019ape and SN 2020oce, representing 1\% of the sample) are featureless ellipticals with no visible signs of star formation.  This could indicate that these events are not actually CC~SNe, that some CC~SNe have long progenitor lifetimes, or that even classical ellipticals contain some residual star formation \citep{Hakobyan+2008,Hakobyan+2012,Graham+2012,Irani+2019}.}  These (and other candidate non-star-forming SN hosts found by ZTF) will be discussed in further detail by Irani et al.\ (in prep.).

Very low-luminosity galaxies contribute negligibly to the SN~Ia population (1.0\%, not including ``hostless'' events) but represent a more significant fraction of the CC~SN population  (5.1\%, plus 1.5\% ``hostless'' events that are probably undetected dwarfs). While there are relatively few low-luminosity galaxies in the SDSS spectroscopic sample by raw numbers, this is largely a result of Malmquist biases inherent in redshift measurement.  These galaxies are quite common by volume and are responsible for a significant fraction of cosmic star formation (although far from a majority; \citealt{Brinchmann+2004}).

The SN~Ia population is noticeably skewed toward being hosted within redder and more luminous galaxies than CC~SNe, as expected for a population that largely traces stellar mass (although a component also traces young stars; e.g., \citealt{Mannucci+2006,Sullivan+2006}).  Notably, however, there is no significant difference in the luminosities or colors of SN~II and SN~Ib/c hosts (a Kolmogorov-Smirnov two-sample test gives $p_{\rm KS} = 0.55$ for $m<18.5$ mag and $p_{\rm KS} = 0.08$ at $m<19$ mag), which may suggest that the importance of metallicity-sensitive channels (such as single-star wind stripping) is minor.  While surprising given earlier work on this topic \citep{Boissier+2009,Arcavi+2010,Graur+2017a,Graur+2017b}, this is in agreement with some other recent observational studies \citep{Anderson+2015,Galbany+2016b,Kuncarayakti+2018,Taggart+2020,StevehostsPTF2020}.

As the SDSS-matched sample is relatively small, we cannot rule out smaller differences in luminosity (the difference in mean absolute magnitude is $0.06\pm0.35$) or color.  We have also not yet investigated explosion-site properties of the sample, which are likely to better reflect differences in delay times among young transients \citep{Anderson+2008,Kelly+2012,Galbany+2014,Maund2018,Xiao+2019}.

\section{Summary and Online Catalog}
\label{sec:conclusions}

In this paper we have summarized the status of the ZTF Bright Transient Survey after two years of operation, and illustrated several basic cuts with which to establish a large, high-quality, unbiased, and nearly spectroscopically-complete sample of extragalactic transients.  Using this sample we have provided a preliminary exploration of transient parameter space on timescales of 6--200 days, a new estimate of the CC~SN rate and luminosity function, and constraints on the fraction of star formation in rare environments such as low-mass and red-sequence galaxies.

Our results are based on ZTF alert packet data and TNS reports, both of which are susceptible to occasional errors. Also, the analyses we have employed are based on simplified general techniques, chosen based on applicability to a wide range of SN properties and the need to avoid human intervention.  These results will eventually be superseded by focused papers on all of these topics using additional classifications, improved redshift measurements, analysis of SN subtypes, superior reference template images and forced photometry, light-curve modeling, host SED fitting, and many other enhancements.  However, we emphasize that the analysis presented here can be updated continuously in real time using public data as the sample continues to expand, and we invite the community to explore the properties of SN parameter space using our public data releases.

To this end, we have created a new web resource summarizing the key properties of our sample and provided an interactive interface to explore it in detail.  Titled the BTS Sample Explorer, it provides a sortable web table containing the time and magnitude of peak light for each transient, classifications and redshifts, timescales measured from the light curve, extinction and luminosity measurements, and host-galaxy photometric properties.  P48 stamp images, light-curve plots, and colorized Pan-STARRS images of the field and host galaxy are also provided.  An alternative viewing mode allows instant collages of light curves, Pan-STARRS cutouts, or both, for public presentation or visual data exploration.  The data table can be downloaded as a .csv file for offline exploration.  All of these resources are updated nightly to add new transient discoveries, classifications, and measurements.  This resource can be accessed at \url{https://sites.astro.caltech.edu/ztf/bts/explorer.php}.

Looking further to the future, while the SN catalog we present here is by far the largest of its type to date, it is clear that even larger-scale efforts will be required to fully address many key science areas: for example, unbiased population studies of uncommon transients (of which we have only a few examples so far) and at the faint end of the luminosity function for common transients.  Several additional years of ZTF operations (with improved cadence and scheduling software, as planned for the second phase of ZTF which is scheduled to begin in October 2020) will be key to this: BTS has been operational for only two years so far, with much of the first year devoted to reference-building.  Combining ZTF with other surveys will also be of significant benefit: close to half of our sample could not be fully utilized due (in part) to light-curve gaps, but data from telescopes at other sites may be able to fill these gaps. 

Key to the success of BTS so far has been the availability of dedicated robotic spectrographs, especially the SED Machine at Palomar.  The commissioning of more such facilities, alongside new high-thoroughput multi-channel spectrographs on existing telescopes, would highly-complete spectroscopic coverage of the transient sky to be extended to greater depths  and larger sky areas.  Several projects of this type are in development, including a proposed SED Machine clone at Kitt Peak, the New Robotic Telescope on La Palma \citep{NRT}, the Next Generation Palomar Spectrograph \citep{NGPS}, and the Son of X-shooter at La Silla \citep{SOXS}.  A concerted, organized effort by these facilities could easily increase the size of highly-complete samples similar to BTS by an order of magnitude or more by the middle of the decade.  This would allow for unambiguous rate and luminosity measurements and strong progenitor constraints for virtually all currently-known classes of transients, discoveries of (or constraints on) even extremely rare and/or fast-evolving events, and greatly advance our understanding of the explosive universe.

{Even with these new generations of spectrographs, the larger but fainter populations of transients found by the Vera Rubin Observatory will pose particular challenges for obtaining similarly complete samples from the transients found by that facility, and heavy reliance on photometric classification will be unavoidable.  Against this backdrop, bright-end surveys supported by large-scale spectroscopic classification efforts (such as ZTF/BTS) remain critical: the wealth of photometric training data and improved rate measurements will aid in developing reliable photometric classification tools and motivating follow-up strategies for the LSST era.}

\acknowledgments

We would like to acknowledge the contributions of the worldwide community of observational astronomers in obtaining and reporting classification spectra of transients within this sample.  We specifically acknowledge observing and classification contributions from Yuhan Yao, Viraj Karambelkhar, Igor Andreoni, and Alison Dugas (Palomar/Keck), WeiKang Zheng, Kishore Patra, and Andrew Hoffman (Lick), Brigitta Spi\H{o}cz, Zach Golkhou, Keaton Bell, and James Davenport (APO).  We also acknowledge the contributions of BTS and other ZTF alert-stream scanners, including Raphael Baer-Way, Teagan Chapman, Matt Chu, Asia deGraw, Suhail Dhawan, Alison Dugas, Nachiket Girish, Samantha Goldwasser, Andrew Hoffman, Connor Jennings, Evelyn Liu, Emily Ma, Emma McGinness, Yukei Murakami, Derek Perera, Druv Punjabi, James Sunseri, Abel Yagubyan, and Erez Zimmerman.    We thank Christopher Cannella and Ashot Bagdasaryan for developing and managing the GROWTH Marshal, and Peter Nugent and Eran Ofek for useful comments on this manuscript.  We also thank Suvi Gezari and Sjoert van Velzen for discussions on the nature of AT2019cmw and AT2019fdr.  We thank the anonymous referee for useful comments.

Based on observations obtained with the Samuel Oschin 48-inch Telescope and the 60-inch Telescope at the Palomar Observatory as part of the Zwicky Transient Facility project. ZTF is supported by the National Science Foundation (NSF) under grant AST-1440341 and a collaboration including Caltech, IPAC, the Weizmann Institute for Science, the Oskar Klein Center at Stockholm University, the University of Maryland, the University of Washington, Deutsches Elektronen-Synchrotron and Humboldt University, Los Alamos National Laboratories, the TANGO Consortium of Taiwan, the University of Wisconsin at Milwaukee, and the Lawrence Berkeley National Laboratory. Operations are conducted by COO, IPAC, and UW.  The SED Machine is based upon work supported by the NSF under grant 1106171.  
The Liverpool Telescope is operated on the island of La Palma by Liverpool John Moores University in the Spanish Observatorio del Roque de los Muchachos of the Instituto de Astrofisica de Canarias with financial support from the UK Science and Technology Facilities Council.
Research at Lick Observatory is partially supported by a generous gift from Google.

Some of the work associated with this paper was carried out at the Aspen Center for Physics (ACP).  The ACP is supported by NSF grant PHY-1607611.  This work was partially supported by a grant from the Simons Foundation.

This work was supported by the GROWTH (Global Relay of Observatories Watching Transients Happen) project funded by the NSF under PIRE grant 1545949. GROWTH is a collaborative project among California Institute of Technology (USA), University of Maryland College Park (USA), University of Wisconsin Milwaukee (USA), Texas Tech University (USA), San Diego State University (USA), University of Washington (USA), Los Alamos National Laboratory (USA), Tokyo Institute of Technology (Japan), National Central University (Taiwan), Indian Institute of Astrophysics (India), Indian Institute of Technology Bombay (India), Weizmann Institute of Science (Israel), The Oskar Klein Centre at Stockholm University (Sweden), Humboldt University (Germany), Liverpool John Moores University (UK), and University of Sydney (Australia).

A.A.M.~is funded by the LSST Corporation, the
Brinson Foundation, and the Moore Foundation in support of the LSSTC Data
Science Fellowship Program; he also receives support as a CIERA Fellow by the
CIERA Postdoctoral Fellowship Program (Center for Interdisciplinary
Exploration and Research in Astrophysics, Northwestern University).
A.G.-Y.’s research is supported by the EU via ERC grant No. 725161, the ISF GW excellence center, an IMOS space infrastructure grant and BSF/Transformative and GIF grants, as well as The Benoziyo Endowment Fund for the Advancement of Science, the Deloro Institute for Advanced Research in Space and Optics, The Veronika A. Rabl Physics Discretionary Fund, Paul and Tina Gardner, Yeda-Sela and the WIS-CIT joint research grant;  A.G.-Y. is the recipient of the Helen and Martin Kimmel Award for Innovative Investigation.
Y.-L.K. has received funding from the European Research Council (ERC) under the European Union's Horizon 2020 research and innovation program (grant agreement No. 759194 - USNAC). A.V.F.'s group is grateful for funding from the TABASGO Foundation, the Christopher J. Redlich Fund, and the Miller Institute for Basic Research in Science (U.C. Berkeley).  M.L.G. acknowledges support from the DiRAC Institute in the Department of Astronomy at the University of Washington. The DiRAC Institute is supported through generous gifts from the Charles and Lisa Simonyi Fund for Arts and Sciences, and the Washington Research Foundation.

Funding for the SDSS and SDSS-II has been provided by the Alfred P. Sloan Foundation, the Participating Institutions, the NSF, the U.S. Department of Energy, the National Aeronautics and Space Administration (NASA), the Japanese Monbukagakusho, the Max Planck Society, and the Higher Education Funding Council for England.     The SDSS is managed by the Astrophysical Research Consortium for the Participating Institutions. The Participating Institutions are the American Museum of Natural History, Astrophysical Institute Potsdam, University of Basel, University of Cambridge, Case Western Reserve University, University of Chicago, Drexel University, Fermilab, the Institute for Advanced Study, the Japan Participation Group, Johns Hopkins University, the Joint Institute for Nuclear Astrophysics, the Kavli Institute for Particle Astrophysics and Cosmology, the Korean Scientist Group, the Chinese Academy of Sciences (LAMOST), Los Alamos National Laboratory, the Max-Planck-Institute for Astronomy (MPIA), the Max-Planck-Institute for Astrophysics (MPA), New Mexico State University, Ohio State University, University of Pittsburgh, University of Portsmouth, Princeton University, the United States Naval Observatory, and the University of Washington.

The Pan-STARRS1 Surveys (PS1) have been made possible through contributions of the Institute for Astronomy, the University of Hawaii, the Pan-STARRS Project Office, the Max-Planck Society and its participating institutes, the Max Planck Institute for Astronomy, Heidelberg and the Max Planck Institute for Extraterrestrial Physics, Garching, The Johns Hopkins University, Durham University, the University of Edinburgh, Queen's University Belfast, the Harvard-Smithsonian Center for Astrophysics, the Las Cumbres Observatory Global Telescope Network Incorporated, the National Central University of Taiwan, the Space Telescope Science Institute, NASA under grant NNX08AR22G issued through the Planetary Science Division of the NASA Science Mission Directorate, the NSF under grant  AST-1238877, the University of Maryland, and Eotvos Lorand University (ELTE).

This research has made use of the NASA/IPAC Extragalactic Database (NED),
which is operated by the Jet Propulsion Laboratory, California Institute of Technology, under contract with NASA.

Some of the data that contributed to this paper were obtained at the W. M. Keck Observatory, which is operated as a scientific partnership among the California Institute of Technology, the University of California, and NASA. The Observatory was made possible by the generous financial support of the W. M. Keck Foundation. The authors wish to recognize and acknowledge the very significant cultural role and reverence that the summit of Maunakea has always had within the indigenous Hawaiian community.

\facilities{PO:1.2m, PO:1.5m, Liverpool:2m, ARC, Shane, Hale, Keck:I (LRIS)}

\bibliographystyle{aasjournal}

\appendix

\section{Filter Details}
\label{sec:filterdetail}

The 2019 version of the Bright Transient Survey filter is designed to pass all genuine transients that would also pass the original filter while reducing the false-positive rate by an order of magnitude.  In detail, requirements of the filter are as follows.

\begin{itemize}
\item The alert must have \texttt{magpsf} $<19$, or a detection in the history with \texttt{magpsf} $< 19$ in the last 18 hr.
\item The subtraction must be positive (\texttt{isdiffpos} $= t$ or \texttt{isdiffpos} $=1$).
\item The location must be outside the Galactic plane: $|b|$ $>$ 7$\degree$.
\item The alert must have \texttt{rbscore}$>0.2$.  If close to a bright catalog object this is increased  to \texttt{rbscore}$> 0.3$ if a $m<17$ mag source is within 1$\arcsec$ and to \texttt{rbscore} $> 0.45$ if a $m<15.5$ mag source is within 1.5$\arcsec$.  (The source magnitude $m$ can be in any filter and can be from PS1 or Gaia.)
\item The alert must have \texttt{drb} $> 0.1$.
\item The alert must not be within \texttt{distpsnr} $<$ 2$\arcsec$ of a high-probability PS1 star (\texttt{sgscore} $>$ 0.76).  It must also not be within 0.5$\arcsec$ of a bright PS1 object with uncertain stellarity (\texttt{sgscore} $=$ 0.5 and $m<17$ mag in any PS1 filter), or within 1.0$\arcsec$ of a very red PS1 source ($r-i>3$ or $r-z>3$ mag, with \texttt{sgscore} $>0.2$).
\item The alert must not be close to a bright potential star among any of the three PS1 sources in the packet.  The exclusion radius depends on the star's magnitude and \texttt{sgscore}.  It is \texttt{distpsnr}$< 20\arcsec$ for stars with $r<15$ or $i<14.5$ mag and sgscore $> 0.8$, and for stars with $r<12$ or $i<11.5$ mag and sgscore $>$ 0.49.  It is \texttt{distpsnr} $< 10\arcsec$ for stars with $z<14.0$ mag and sgscore $>0.8$, or $z<11.5$ mag and sgscore $>0.49$.  It is \texttt{distpsnr} $<5\arcsec$ for stars with $r<15$ or $i<14.5$ mag and \texttt{sgscore} $>0.49$. It is \texttt{distpsnr} $<2.5\arcsec$ for stars with $z<14$ mag and \texttt{sgscore} $>0.49$.  It is \texttt{distpsnr} $<1.1\arcsec$ for stars with $r<16.5$ or $i<16.0$ mag and sgscore $>0.49$.  It is \texttt{distpsnr} $<0.9\arcsec$ for stars with $z<15.5$ mag.  The most restrictive (largest exclusion radius) is always used.  While an improvement over our 2018 filter, this criterion was found to occasionally reject real SNe and has been further revised in mid-2020.
\item To remove moving objects, there must be another alert at the same location in the history more than 0.02 days prior.
\item The alert must also not have a cross-match in the minor planet catalog within \texttt{ssdistnr} $<15\arcsec$.
\item The alert must not be variable, as determined by the presence of a coincident counterpart and a first detection well before the alert (generally $>90$ days, based on $dt$ = \texttt{jd} $-$ \texttt{jdstarthist}).  The counterpart matching radius is \texttt{distnr} $<0.4\arcsec$ for \texttt{magnr} $<19.5$ mag,  \texttt{distnr} $<0.8\arcsec$ for \texttt{magnr} $<17.5$ mag, \texttt{distnr} $<1.2\arcsec$ for \texttt{magnr} $<15.5$ mag, and \texttt{distnr} $<9.5\arcsec$ for \texttt{magnr} $<$ \texttt{magpsf} $-1$; it is \texttt{neargaia} $<0.35\arcsec$ for \texttt{maggaia}$<17$ mag, and \texttt{neargaia} $<0.20\arcsec$ for \texttt{magggaia} $<18$ mag.  Sources with  \texttt{neargaia} $<0.35\arcsec$ and \texttt{maggaia} $<19$ mag are also excluded but only if the alert is $m>18.5$ mag and $dt>300$ days.   A historical alert can be generated for spurious regions (especially near galaxy centers where bad subtractions are common) so caution is necessary in applying this filter: the criteria based on the reference catalog are only applied if the light curve is not at a local maximum and there are already several $m<19$ mag detections in the history (an indicator that the source has passed the filter before and not been saved).
\end{itemize}

The above summary is slightly simplified and the associated changes were not all made simultaneously.  Prior to June 2019, selection was performed using the basic filter described by \citep{Fremling2020}; after June 2020, the filter above was updated to decrease the exclusion radius around bright stars but cut more strictly on \texttt{drb}.  This further-improved filter passed all of the TNS-cataloged transients that we missed on account of star-galaxy confusion (\S \ref{sec:completeness}) with the exception of SN 2019gcc, a nuclear SN Ia which had a few detections several months prior to explosion that could be due to activity from a coincident weak AGN.   Python code for all three versions (2018, 2019, and 2020) is available online \footnote{\url{https://github.com/dperley/ztf-bts-filters} {\citep{BTSfiltercode}}}.

Note that \texttt{drb} became available in alert packets only in summer 2019, and asteroid and Gaia matching were also not available for packets early in the survey.  When running the filter retroactively (for the purposes of our quality cut and verification checks) any criteria associated with missing fields are not applied.

\section{Cross-Match Associations for Purity Filter}
\label{sec:xmatch}

Any transient candidate saved to the program which passes the quality cuts and which has a ``supernova-like'' timescale will be included in our sample.  To avoid excluding any potentially very fast or very slow transients, we also pass all transients with credible host-galaxy associations even if their timescale is not supernova-like.  We perform two host-galaxy association checks using cross-matches of different catalogs.  

\subsection{Pan-STARRS cross-match}
\label{sec:xmatchps1}

\begin{figure*}
\centering
\includegraphics[width=18cm]{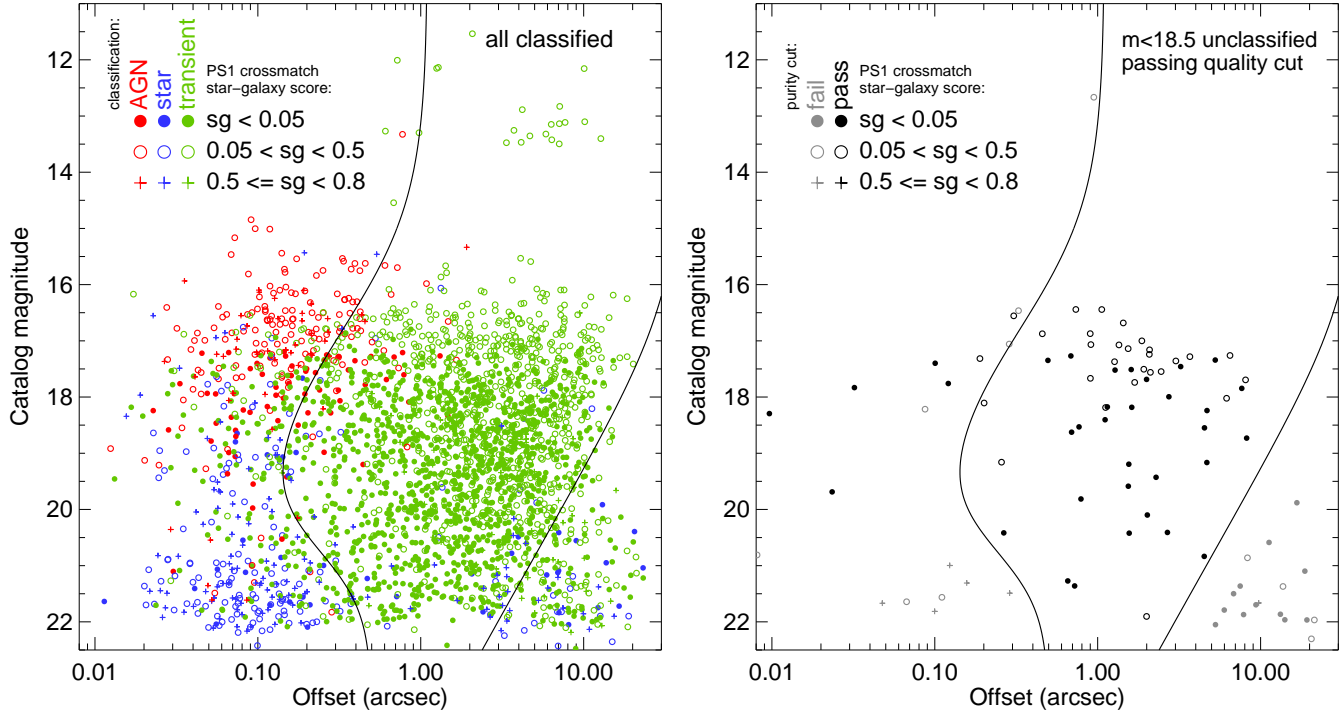}
\caption{Angular offset versus magnitude for Pan-STARRS catalog cross-matches.  The left panel shows classified events saved to the program, color-coded by type (genuine transients, stars/CVs, or AGNs).  The symbol indicates the star-galaxy score (\citealt{Tachibana2018}; higher values are more star-like).  The right panel shows unclassified transients passing our quality cut (\S\,\ref{sec:quality}).  Events between the two lines automatically pass the purity cut as long as {\tt sgscore} $<0.5$.  Events further to the left automatically pass if {\tt sgscore} $<0.05$.  Events in the bottom-right corner are likely to be chance associations.}
\label{fig:xmatchps1}
\end{figure*}

The first check involves the Pan-STARRS 1 catalog.  The nearest three Pan-STARRS matches, with \texttt{sgscore} values, are located in the Avro packet data; we generally take the nearest cross-match of the three, although if the nearest source has \texttt{sgscore} $>$ 0.75 we will use one of the other two sources if it has \texttt{sgscore} $<$ 0.75 and is within 5$\arcsec$.  A plot of the magnitude versus offset of PS1 matches is shown in Figure 10, for both classified and unclassified events \emph{including} all non-transient false-positives saved to the program.  We only show cross-matches with \texttt{sgscore} $<0.8$.  For moderate offsets the plot is overwhelmingly dominated by genuine transients.  For very small offsets the transient population is contaminated by two other populations: AGNs at bright magnitudes and CVs at faint magnitudes; these are generally cases where \texttt{sgscore} has miscategorized the source.  It is also contaminated by CVs at very large offsets and faint magnitudes  (in this case due to chance occurrence near a galaxy).

The diagram is subdivided by lines into three regions: coincident cross-matches (at left), offset but highly-probable cross-matches (center), and likely-spurious cross-matches (lower right).  The equation defining a coincident match is
\begin{equation}
\theta < 0.1+50\times10^{-0.2(m-15.8)}(0.3+0.7(1+e^{15.8-m})^{-1}). \end{equation}
\noindent
The equation defining a spurious match is
\begin{equation}
\theta > 0.1+0.4(1+0.5\times10^{-0.8(m-21.5)})^{-1}+(1+10^{+0.2(m-15.5))})^{-1}.
\end{equation}
\noindent
Here, $\theta$ is the offset in arcsec and $m$ is the catalog magnitude of the candidate cross-match ($r$ magnitude when available but another band is used if no $r$ photometry exists).  
The exact form of these equations is arbitrary and was chosen largely using trial and error in order to avoid as much of the non-transient populations as possible while still including the vast majority of real transients.  The slope term of $-$0.2 is motivated by the assumption that galaxies on average have constant surface brightness on the sky (in general, dimmer galaxies are proportionally smaller), while the increasing offset at faint magnitudes reflects increasing positional measurement uncertainties for faint cross-matches.

Transient candidates within the middle region of the diagram (offset, probable matches) pass the purity cut if the cross-match has \texttt{sgscore} $<$ 0.5.  Transient candidates in the left of the diagram (coincident matches consistent with no offset) pass if the cross-match has \texttt{sgscore} $<$ 0.05.  Transients in the bottom right do not automatically pass the purity cut.

\subsection{Lasair cross-match}
\label{sec:xmatchlasair}

\begin{figure*}
\centering
\includegraphics[width=18cm]{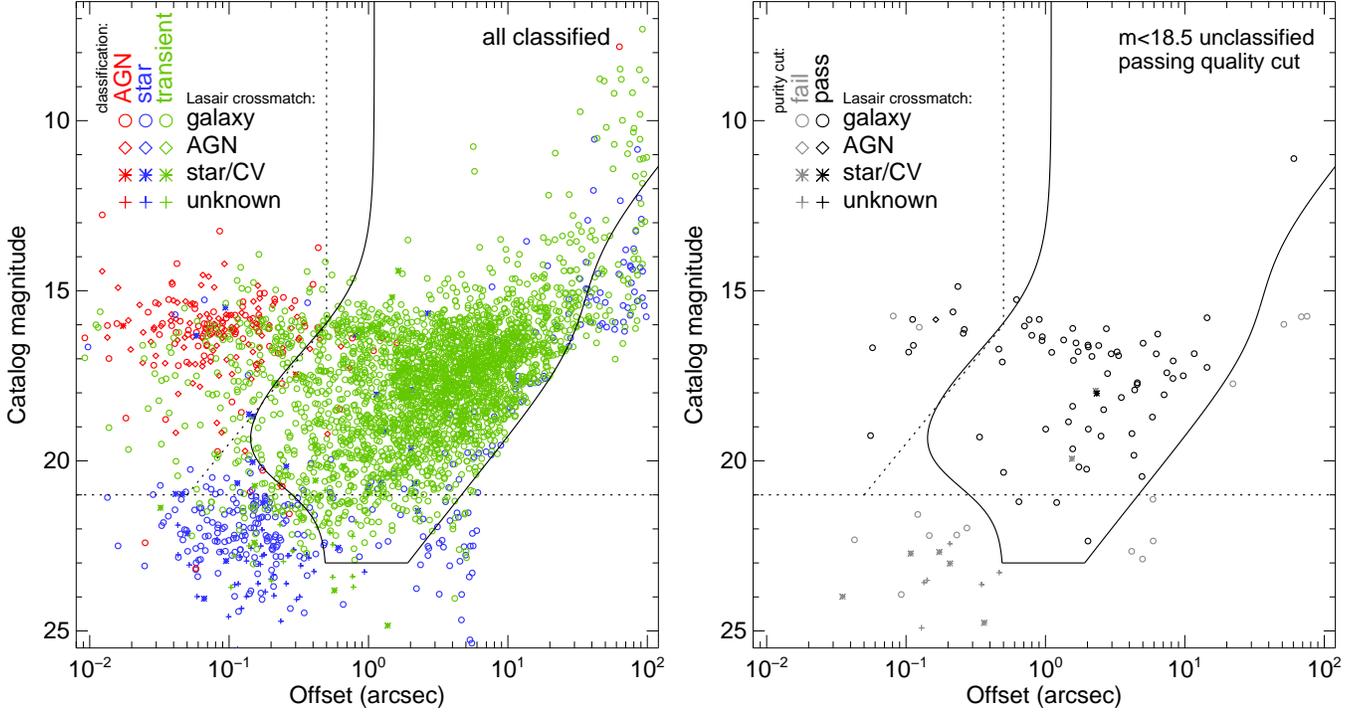}
\caption{Angular offset versus magnitude for the most likely cross-matched source provided by Lasair.  Panels and colors are as in Figure~\ref{fig:xmatchps1}, but here the symbol indicates the type designation in the Lasair catalog.  The lines are the same as in Figure~\ref{fig:xmatchps1}.  In this case, only events in the middle region associated with galaxies or ``faint/unknown'' automatically pass the purity cut.}
\label{fig:xmatchlasair}
\end{figure*}

The Pan-STARRS matches within the Avro packets only include the nearest three sources within $30\arcsec$, making it not particularly useful for transients in nearby and large galaxies.  A useful \texttt{sgscore} value is also not always available (\S\,\ref{sec:filter}).  We therefore also perform a second cross-match, relying on the cross-match tool provided by Lasair which searches for galaxy catalog associations out to much larger radii.  For each candidate in our program, the list of potential cross-matches is obtained from this service, along with their $g$ and $r$ magnitudes and ``type'' (which can be ``star'', ``galaxy'', ``agn'', and occasionally other types such as ``cv''.)  We reassign ``agn'' types to ``galaxy'' if the offset is more than $1\arcsec$.

A plot of offsets versus magnitudes from Lasair is shown in Figure~\ref{fig:xmatchlasair}.  The general appearance is quite similar to the previous figure but extending to larger offsets and brighter galaxies.  The spurious region of the diagram is mostly empty, since these are generally not cross-matched by Lasair in the first place.

We use the same equations to identify physical, non-coincident cross-matches as in \S\,\ref{sec:xmatchps1}, although we also add a magnitude cut of $m<23$.  Transients with cross-matches in the central region of the diagram bounded by these three lines, for which the cross-match is a ``galaxy,'' automatically pass the purity cut.

\section{Choosing Between Multiple Host Cross-Matches}
\label{sec:xmatchprob}

It is frequently the case that there are several candidate host-galaxy cross-matches.  In these cases it is important to determine which (if any) is the most credible host galaxy.

We use a simple least-likelihood method by calculating the probability that a given position, had it been randomly chosen across the entire sky, would be located as close or closer to a galaxy as bright or brighter than the host galaxy candidate under consideration.  The general equation for this probability is

\begin{equation}
p = 1 - {\rm exp}(-\pi \theta^2 \rho),
\end{equation}
\noindent
where $\theta$ is the angular offset and $\rho$ is the sky density of galaxies at least as bright (in apparent magnitude) as the putative host in the given filter band.  

Because our cross-matching is automatic and must deal with shredded galaxies (which may have components very close to the transient), we employ a few approximations and modifications to this basic approach.  We use a simplified single power law of $\rho = 220 \times 10^{0.55(m-18)}$ deg$^{-2}$, which provides a reasonable approximation to the $r$-band number counts in \cite{Yasuda+2001} at the bright end.  Since we are only interested in highly probable cross-matches, we can also safely approximate $1-e^{-x}$ as $x$.  Finally, because matches closer than $1\arcsec$ are not meaningful for galaxies within our distance limit (ZTF pixels are $1\arcsec$  in size and few galaxies are smaller than $1\arcsec$ in size), we de-weight cross-matches of order 1$\arcsec$ by substituting $\theta$ with $\theta+1\arcsec$.  Therefore, the actual equation used in practice is

\begin{equation}
p = \pi \left( \frac{\theta_{\prime\prime}+1}{3600} \right)^{2} 220 \times 10^{-0.55(m-18)}.
\end{equation}
\noindent
The galaxy with the lowest $p$-value is chosen as the association.

For the purposes of sample selection, only the relative value of the $p$ is meaningful; chance associations are rejected on a purely empirical basis as described in \S\,\ref{sec:xmatchps1}.  For host-galaxy assignment, we designate the transient as hostless if there is no galaxy within SDSS with a value of $p<0.1$ or, if SDSS is not available, no PS1 galaxy with $p<0.05$.  (The stricter cut for PS1 is due to the higher incidence of spurious sources in this catalog.)  All SDSS host associations used in the host-galaxy analysis of this paper (\S\,\ref{fig:hostgalaxies}) were manually vetted and the host reassigned if the automatically-determined host was assessed to be inaccurate.

\clearpage

\section{Seasonal dependence of classification rate}
\label{sec:season}

\begin{figure*}
\centering
\includegraphics[width=18cm]{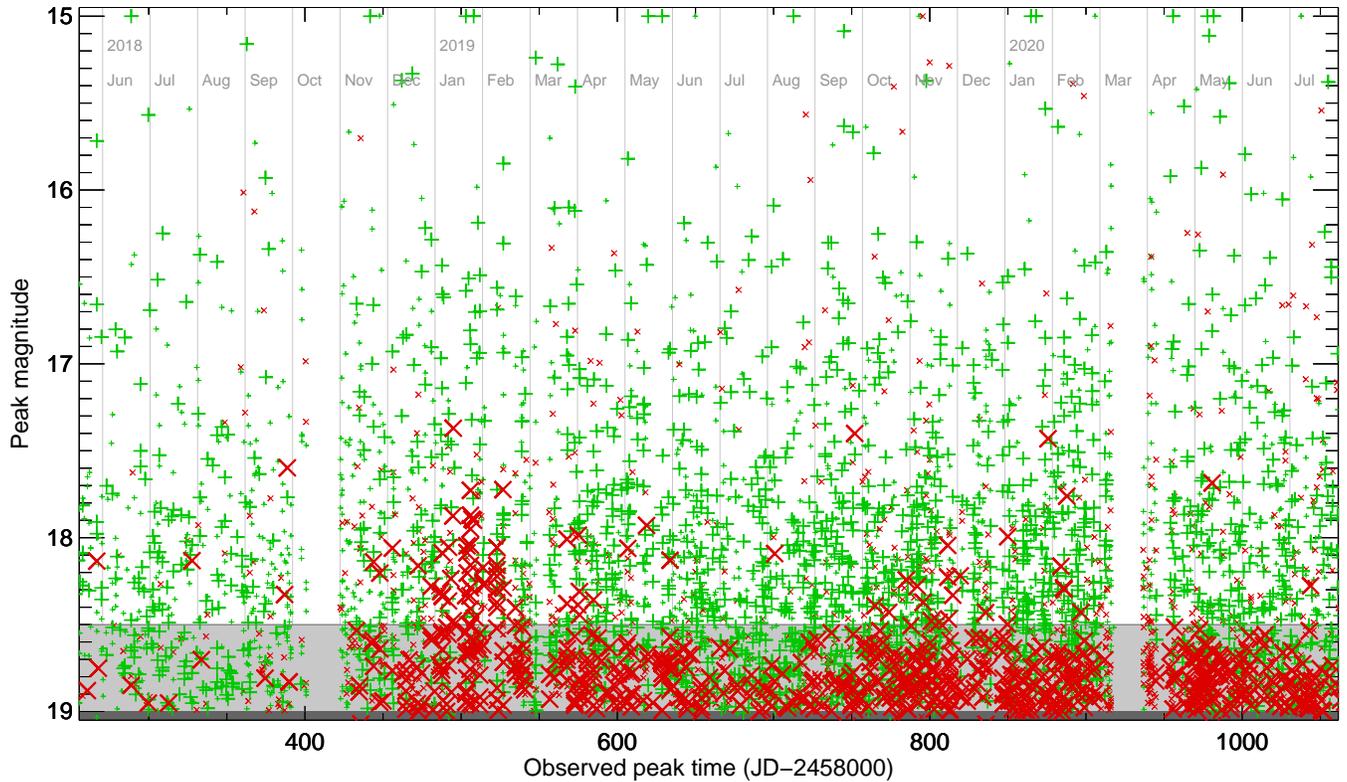}
\caption{Classification success or failure by time and magnitude of observed light-curve peak.  Green ``$+$'' symbols indicate successful classifications and red ``$\times$'' symbols indicate missing classifications.  Large symbols indicate candidate transients which passed the quality and purity cuts; small symbols indicate candidate transients which did not pass these cuts.  The primary determining factor governing classification success (at $m<18.5$ mag) is the impact of weather on spectroscopic follow-up runs: poor conditions affecting classical runs and P60 operations in early 2019 (and to a much less extend late 2019) led to a larger fraction of missing classifications in these periods.  The gap in October 2018 is due to maintenance and the gap in March 2020 is due to extended bad weather.  Transients with $m<15$ mag are fixed at $m=15$ mag in this plot.}
\label{fig:timeplot}
\end{figure*}

In Figure~\ref{fig:timeplot} we plot all candidate transients saved to the program by the time and magnitude of peak, color-coded by whether the event was classified or unclassified.  This shows clearly the seasonal dependence of the success of our spectroscopic follow-up observations: during most of the year (summer and autumn, especially) we are almost 100\% complete to $m<18.5$ mag, but during the winter months when SEDM cannot operate for long stretches and classical follow-up runs may be weathered out, there are occasional periods where significant numbers of brighter transients are also missed.

\clearpage

\section{Reclassifications}
\label{sec:reclassification}

In almost all cases the classification we have used is the most recent classification associated with that object on TNS, although we remove some subtype information because (for spectra reported by us) we are not yet able to uniformly distinguish classical SN Ia subtypes or to separate SN IIP vs. SN IIL.  

In a few cases we have used a different classification from what is reported on TNS --- either because the classification was reported in a public reference other than TNS or because the TNS classification appears to be in error.  We list these in Table~\ref{tab:reclassify}.

\begin{deluxetable}{lllll}
\tabletypesize{\footnotesize}
\tablecolumns{5}
\tablewidth{0pt}
\tablecaption{Non-TNS classifications \label{tab:reclassify}}
\tablehead{
\colhead{ZTF ID} & \colhead{IAU ID} & \colhead{Classification} & \colhead{Redshift} & \colhead{Reference/note}
}
\startdata
ZTF18aamfrvy & SN2018ahe & --    & 0.01564  & Added redshift from NED \\
ZTF18aazgfkq & SN2018cmk & --    & 0.025724 & Added redshift from NED \\
ZTF18abcfcoo & AT2018cow & other & 0.014145 & New/peculiar transient class \\
ZTF18actuhrs & SN2018evt & SN Ia-CSM & --   & This work \\
ZTF19aadnwvc & AT2019ye  & SN Ia & 0.077    & ATEL12426 \\
ZTF19aagqkrq & AT2019ahd & ILRT  & --       & This work \\
ZTF19aaniqrr & AT2019cmw & other & 0.519    & New/peculiar transient class; paper in prep. \\
ZTF19aaplpaa & SN2019cxx & --    & 0.025    & Added redshift from NED \\
ZTF19aatubsj & SN2019fdr & none  & --       & Possibly an AGN/NLSy1 based on late-time spectra \\
ZTF19aatevrp & SN2019dke & --    & 0.010637 & Added redshift from NED \\
ZTF19aavxfib & AT2019gte & none  & --       & TDE classification is uncertain \\
ZTF19acdsqir & SN2019sxd & --    & 0.066    & Added redshift from NED \\
ZTF19acnfsij & ST2019uiz & nova  & M31      & ATEL 13317 \\
ZTF19acoaiub & AT2019udc & ILRT  & --       & This work \\
ZTF19adakuos & AT2019wvf & nova  & M31      & ATEL 13384 \\
ZTF20aaertpj & AT2020pv  & SN Ib & 0.02875  & GCN 26703 \\
ZTF20aaeuxqk & SN2020ut  & --    & 0.035    & Revised redshift \\
ZTF20aakdppm & AT2020ber & nova  & M31      & Recurrent nova M31N 1926-07c \\
ZTF20aatwonv & SN2020euz & --    & 0.0226   & Added redshift from NED \\
ZTF20abijfqq & SN2020nlb & --    & 0.002432 & Added redshift from NED \\
ZTF20abfhyil & SN2020mrf & none  & none     & \begin{tabular}{@{}l@{}}TNS classification based on \\ featureless spectrum; probable CV \end{tabular}\\
\enddata
\tablecomments{An empty field (--) indicates that we retain the existing TNS classification or TNS redshift.}
\end{deluxetable}

\clearpage

\section{The BTS Sample Explorer}
\label{sec:explorer}

To facilitate public use of our transient sample we have developed a web-based interface to display the sample (including objects that are not transients, which fail our cuts, etc.) and further filter it in various ways: by date of maximum light, by classification, by redshift, by peak magnitude (apparent or absolute), and by many other properties.  It also provides P48 postage-stamp images of the transient and/or reference images, three-color PS1 images combined according to the method of \citep{Lupton+2004}, and light-curve plots.

The website front-end is built in basic PHP.  A Python back-end is used to build and update the database; this back-end is the same as the one we have used to calculate timescales, cross-matches, and all other properties discussed in this paper.
The back-end scripts to calculate these properties are executed automatically via \texttt{cron} every 3 hr.  The scripts update the underlying data files (alert data, PS1 FITS images, TNS classification tables, etc.) by downloading from the relevant sources on a regular basis --- daily for recent transients, less often for transients with a long history --- or anytime a new event is saved to the program.  Light-curve plots are also regenerated if new data points appear.
 
User queries execute extremely fast (within 1~s), although if images are requested these can take somewhat longer to load in their entirety for very large queries.

The interface is currently relatively basic: users can select from a range of options and enter start and end values for the purposes of filtering on various properties.  More complex SQL-style queries are not yet possible although this is planned for the future.

Data are normally displayed as a table, but a grid mode can also be selected to specifically display the images (PS1 cutouts, light curves, or both).

Currently available data columns include the ZTF, IAU, and discoverer identifiers, the peak time and magnitude, the coordinates ($\alpha$, $\delta$), the half-peak-to-peak rise and fade times (and the sum of these values, the ``duration''), the classification and redshift, the absolute magnitude of the transient, host absolute magnitude and color, Galactic latitude and Galactic extinction, and sample selection flags.

\begin{figure*}
\centering
\includegraphics[width=16cm]{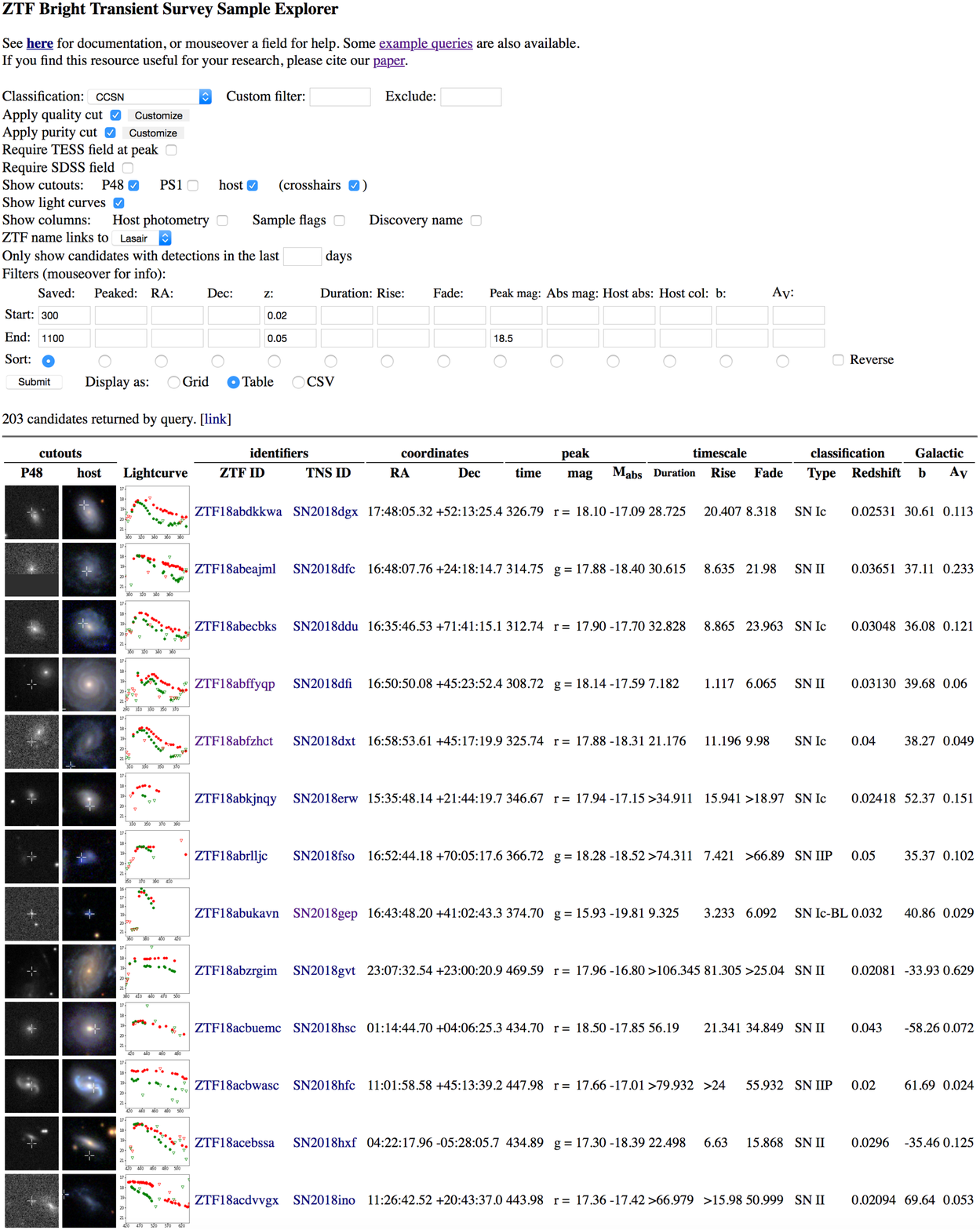}
\caption{Example screenshot from the BTS Sample Explorer.}
\label{fig:screenshot}
\end{figure*}

Further documentation can be found on the Explorer section of the BTS website\footnote{\url{https://sites.astro.caltech.edu/ztf/bts/explorer.php}}.  An example screenshot is presented in Figure~\ref{fig:screenshot}.

\clearpage

\end{document}